\documentclass[twocolumn,aps,prb,amsmath,amssymb,superscriptaddress]{revtex4-2}
\usepackage{graphicx}
\usepackage{dcolumn}
\usepackage{bm}
\usepackage{braket}
\usepackage{hyperref}
\hypersetup{
    colorlinks=true,
    linkcolor=blue,
    urlcolor=blue,
    citecolor=blue,
}
\usepackage[capitalise]{cleveref}
\usepackage[euler]{textgreek}
\usepackage{xcolor}
\usepackage[version=4]{mhchem}
\usepackage{commath}
\usepackage{multirow}

\begin{document}

\title{Probing the Evolution of Electron Spin Wavefunction of NV Center in diamond via Pressure Tuning}

\author{Kin On Ho}
 \thanks{These authors contributed equally to this work.}
\affiliation{
Department of Physics, The Chinese University of Hong Kong, Shatin, New Territories, Hong Kong, China
}
\author{Man Yin Leung}
 \thanks{These authors contributed equally to this work.}
\affiliation{
Department of Physics and the IAS Centre for Quantum Technologies, The Hong Kong University of Science and Technology, Clear Water Bay, Kowloon, Hong Kong, China}
\author{P. Reddy}
 \thanks{These authors contributed equally to this work.}
\affiliation{
Laser Physics Centre, Research School of Physics and Engineering, Australian National University, 2601, Australia
}
\author{Jianyu Xie}
 \thanks{These authors contributed equally to this work.}
\affiliation{
Department of Physics, The Chinese University of Hong Kong, Shatin, New Territories, Hong Kong, China
}
\author{King Cho Wong}
 \thanks{These authors contributed equally to this work.}
\affiliation{
Department of Physics, The Chinese University of Hong Kong, Shatin, New Territories, Hong Kong, China
}
\author{Yaxin Jiang}
\author{Wei Zhang}
\author{King Yau Yip}
\affiliation{
Department of Physics, The Chinese University of Hong Kong, Shatin, New Territories, Hong Kong, China
}
\author{Wai Kuen Leung}
\affiliation{
Department of Physics and the IAS Centre for Quantum Technologies, The Hong Kong University of Science and Technology, Clear Water Bay, Kowloon, Hong Kong, China}
\author{Yiu Yung Pang}
\author{King Yiu Yu}
\affiliation{
Department of Physics, The Chinese University of Hong Kong, Shatin, New Territories, Hong Kong, China
}

\author{Swee K. Goh}
\affiliation{
Department of Physics, The Chinese University of Hong Kong, Shatin, New Territories, Hong Kong, China
}

\affiliation{
Shenzhen Research Institute, The Chinese University of Hong Kong, Shatin, New Territories, Hong Kong, China
}

\author{M. W. Doherty}
 \email{marcus.doherty@anu.edu.au}
\affiliation{
Laser Physics Centre, Research School of Physics and Engineering, Australian National University, 2601, Australia
}

\author{Sen Yang}
 \email{phsyang@ust.hk}
\affiliation{
Department of Physics, The Chinese University of Hong Kong, Shatin, New Territories, Hong Kong, China
}
\affiliation{
Department of Physics and the IAS Centre for Quantum Technologies, The Hong Kong University of Science and Technology, Clear Water Bay, Kowloon, Hong Kong, China}

\date{\today}

\begin{abstract}
Understanding the profile of a qubit's wavefunction is key to its quantum applications. Unlike conducting systems, where a scanning tunneling microscope can be used to probe the electron distribution, there is no direct method for solid-state-defect based qubits in wide-bandgap semiconductors. In this work, we use pressure as a tuning method and a nuclear spin as an atomic scale probe to monitor the hyperfine structure of negatively charged nitrogen vacancy (NV) centers in diamonds under pressure. We present a detailed study on the nearest-neighbor \ce{^13C} hyperfine splitting in the optically detected magnetic resonance (ODMR) spectrum of NV centers at different pressures. By examining the \ce{^13C} hyperfine interaction upon pressurizing, we show that the NV hyperfine parameters have prominent changes, resulting in an increase in the NV electron spin density and rehybridization from $sp^3$ to $sp^2$ bonds. The \textit{ab initio} calculations of strain dependence of the NV center's hyperfine levels are done independently. The theoretical results qualitatively agree well with experimental data without introducing any fitting parameters. Furthermore, this method can be adopted to probe the evolution of wavefunction in other defect systems. This potential capability could play an important role in developing magnetometry and quantum information processing using the defect centers.
\end{abstract}

\maketitle

\section{Introduction}
The study of a qubit's wavefunction is the heart of solving quantum systems, yet the theoretical computation of such wavefunctions is often demanding. Unlike conducting systems whose wavefunctions can be practically probed by a scanning tunneling microscope, the wavefunctions of point defects in wide-bandgap semiconductors cannot be directly measured, since the defects are highly confined in the insulating host material. Hence, understanding the wavefunctions of such defects is also experimentally challenging. It was proposed that the responses of a solid-state defect's wavefunction to thermal expansion or mechanical stress can be studied via measuring the hyperfine interactions between the defect and its nearby nuclear spins \cite{Barson2017, Doherty2014P, Barson2019, soshenko2020Temperature, Jarmola2020Robust, Wang2022Characterizing, Tang2022First}. Thermal expansion is typically a weak effect and noticeable changes have not been observed until recently \cite{Barson2019, soshenko2020Temperature, Jarmola2020Robust, Wang2022Characterizing, Tang2022First}. On the other hand, stress is a much stronger lattice tuner than temperature. In this work, we demonstrate the measurement of stress-dependent hyperfine interaction in a point defect system for the first time and an accord between theoretical and experimental results without introducing any fitting parameters.

The point defect we investigated is the negatively charged nitrogen vacancy (NV) center in diamond, a color center consisting of a substitutional nitrogen atom and an adjacent carbon vacancy. The NV center has unprecedented potential in the fields of quantum metrology \cite{Yip2019, Ho2020, Lesik2019, Hsieh2019, Doherty2014P, Kucsko2013, Kucsko2013, Dolde2011, Nusran2018, Broadway2019, Joshi2019, Schlussel2018, Thiel2016, Steele2017, Neumann2013}, information \cite{Maurer2012, Bradley2019, Cooper2020}, and communication \cite{Yang2016, Hensen2015, Neumann2008}. In particular, it is a promising sensor with superior spatial resolution and sensitivity, and its electron spin resonance (ESR) can be readily measured via the optically detected magnetic resonance (ODMR) method due to the spin-state-dependent fluorescence rate. The transitions between the triplet ground states $\Ket{m_{s}} = \Ket{0}$ and $\Ket{m_{s}} = \Ket{\pm 1}$ are found to be approximately 2870 MHz at ambient conditions \cite{Doherty2013}. Several pioneer works have revealed the stress susceptibility of the NV center \cite{Gali2019Ab, Doherty2012Theory, Falk2014Electrically, Udvarhelyi2018Spin, Maze2011Properties, Doherty2011The, Doherty2014Temperature, Ivady2014Pressure, Kobayashi1993High, Deng2014New, Udvarhelyi2018Ab}, but the fundamental origin of this response is not completely clear yet, for instance how the spin density and hybridization ratio of the NV center’s orbitals may change under stress. Precision quantum sensing and high-fidelity quantum information processing operations require well-characterized susceptibilities and an accurate understanding of the physical origin, and stress is one of the least explored parameters.

To probe the evolution of electron spin wavefunction of an NV center, one has to use the nearest-neighbor nuclear spins to have enough overlap with the wavefunction of the NV unpaired electrons. In a natural diamond, a small natural abundance (1.1\%) of carbon atoms exists as \ce{^13C} with nuclear spin $I=$1/2. If one of the nearest carbon atoms is \ce{^13C} instead of \ce{^12C}, the NV electron spin couples with it via strong magnetic dipole-dipole interaction and Fermi contact interaction \cite{Loubser1978}, and the hyperfine coupling is about 127 MHz at ambient conditions \cite{Simanovskaia2013, Barson2019, Felton2009, Nizovtsev2010, Mizuochi2009, Baranov2011, Gali2008}. This is two orders of magnitude stronger than another common hyperfine coupling with the nearest \ce{^14N} nuclear spin which is around 2 MHz. The nearest \ce{^13C} neighbors can thus be an ideal probe for the change in the NV center wavefunction under stress.

Upon applying hydrostatic pressure, we expect the nuclei surrounding the NV center to change their positions significantly \cite{Barson2017, Doherty2014P}, so the NV electron spin density and orbital hybridization of the unpaired NV spin would change accordingly \cite{Barson2019}. These wavefunction responses are encrypted in the pressure-dependent \ce{^13C} hyperfine parameters of the NV center. In this work, we first outline the theory behind the \ce{^13C} hyperfine interaction and the NV electron spin wavefunction. We then experimentally study the resonances in the ODMR spectrum associated with the nearest-neighbor \ce{^13C} hyperfine interaction under hydrostatic pressure. This is an exciting attempt to measure the changes in the atomic level of the NV center via external perturbations.

\section{Model}
Using the basis defined as $\Ket{m_{s}, m_{I}} = \{\Ket{1,\frac{1}{2}}, \Ket{1,-\frac{1}{2}}, \Ket{0,\frac{1}{2}}, \Ket{0,-\frac{1}{2}}, \Ket{-1,\frac{1}{2}}, \Ket{-1,-\frac{1}{2}}\}$, the NV+\ce{^{13}C} Hamiltonian is given by \cite{Nizovtsev2010}

\begin{equation}
H=\left(
\begin{array}{cccccc}
 a+\frac{D}{3} & \sqrt{2} d^* & d^* & c^* & 0 & 0 \\
 \sqrt{2} d & \frac{D}{3}-a & b & -d^* & 0 & 0 \\
 d & b & -\frac{2 D}{3} & 0 & d^* & c^* \\
 c & -d & 0 & -\frac{2 D}{3} & b & -d^* \\
 0 & 0 & d & b & \frac{D}{3}-a & -\sqrt{2} d^* \\
 0 & 0 & c & -d & -\sqrt{2} d & a+\frac{D}{3} \\
\end{array}
\right).
\label{Hamil}
\end{equation}

In a coordinate system where the NV and \ce{^{13}C} reference frames coincide in the $x$ axes \cite{Nizovtsev2010I}, the matrix elements can be expressed as \cite{SI}:

\begin{equation}
a=\frac{A_{zz}}{2},
b=\frac{A_{xx}+A_{yy}}{2 \sqrt{2}},
c=\frac{A_{xx}-A_{yy}}{2 \sqrt{2}},
d=i\frac{A_{yz}}{2 \sqrt{2}}.
\label{abcd}
\end{equation}

\begin{align}
A_{xx}&=A_{\bot}, \theta = 109.47\textsuperscript{o},\nonumber\\
A_{yy}&=A_{\parallel} \sin ^2\theta+A_{\bot} \cos ^2\theta,\nonumber\\
A_{zz}&=A_{\parallel} \cos ^2\theta+A_{\bot} \sin ^2\theta,\nonumber\\
A_{yz}&=A_{zy}=(A_{\parallel}-A_{\bot}) \sin \theta \cos \theta.
\label{Axyz}
\end{align}

Effectively, there are only three variables: $A_{\parallel}$, $A_{\perp}$ and $D$. This 6x6 matrix has three doubly degenerate roots which can be analytically solved by the Cardano formulas (details in \cite{Nizovtsev2010, SI}). The transitions from these three NV+\ce{^{13}C} eigenvalues (energy levels) can be readily measured from the dips in our ODMR spectra. The calculations can be generalized to the cases under pressure, implying that we can solve $A_{\parallel}(P)$ and $A_{\perp}(P)$ from $\Delta_{hf}(P)$, $\delta_{hf}(P)$, and $D(P)$ under pressure $P$ measured by NV centers ($dD/dP = 1.49$ MHz/kbar \cite{Ho2020}).

Using hyperfine parameters $A_{\parallel}(P)$ and $A_{\perp}(P)$ determined from ODMR parameters $\Delta_{hf}(P)$ and $\delta_{hf}(P)$, we can compute the Fermi contact term $f(P)$ and the dipole term $d(P)$, thus the electron spin density $\eta(P)$ and the hybridization of $p$ orbitals $\abs{c_{p}(P)}^{2}$. They are related as \cite{Barson2019, Loubser1978, He1993, Morton1978}

\begin{align}
A_{\|}(P) &= f(P) + 2d(P),\\
A_{\perp}(P) &= f(P) - d(P).
\label{AFermi}
\end{align}

\begin{align}
f(P) &= 3777 \times \left(1 - \abs{c_{p}(P)}^{2}\right) \eta(P) \text{ MHz},\\
d(P) &= 107.4 \times \abs{c_{p}(P)}^{2} \eta(P) \text{ MHz},
\label{fdTerm}
\end{align}
where $\eta$ is the electron spin density at the nucleus and $\abs{c_{p}}^{2}$ is the hybridization of $p$ orbitals. These relations show that measuring $A_{\|}$ and $A_{\perp}$ to obtain $f$ and $d$ allows for the determination of the pressure-dependent changes in the electron spin density $\eta$ and the NV orbital hybridization $\abs{c_{s}}^{2}/\abs{c_{p}}^{2}$, where $\abs{c_{s}}^{2}$ is the hybridization of $s$ orbitals. Here, it is assumed that the contribution to the total molecular orbitals (MOs) from an atomic orbital at the nuclear spin is a hybrid orbital $\psi$ which is a linear combination of $s$ ($\phi_{s}$) and $p$ orbitals ($\phi_{p}$) \cite{Loubser1978}, and $\psi$ satisfies

\begin{align}
\psi &= c_{s} \phi_{s} + c_{p} \phi_{p},\nonumber\\
1 &= \abs{c_{s}}^{2} + \abs{c_{p}}^{2}.
\label{orbitals}
\end{align}

\section{DFT. Calculation}
Independent \textit{ab intio} calculations were performed to derive the NV+\ce{^{13}C} hyperfine structure under applied stress for comparison to experimental findings \cite{SI}. These calculations were performed using the Vienna Ab initio Simulation Package (VASP). We studied a 512 atom supercell containing an NV center with a \ce{^{13}C} nearest neighbor. All calculations used a 600 eV plane-wave cutoff energy and the Perdew-Burke-Ernzerhof (PBE) functional. The hyperfine interaction was evaluated in the NV ground state using the in-built hyperfine routine from VASP.

We calculated zero-field hyperfine energies that are in agreement with the accepted theoretical values: $\Delta_{hf}(0) = 127.604 \; \mathrm{MHz}$ and $\delta_{hf}(0) = 2876.86 \;\mathrm{MHz}$. The dependence of $\Delta_{hf}$ on stress was evaluated as follows:

\begin{align}
\frac{\partial\Delta_{hf}}{\partial\sigma_{ij}} = \begin{bmatrix}
0.0487 & 0.0067 & -0.0063\\
0.0067 & 0.0357 & 0.0003\\
-0.0063 & 0.0003 & 0.0342\\
\end{bmatrix} \mathrm{MHz}/\mathrm{kbar},
\end{align}
where $\sigma_{ij}$ are the stress tensor components and $i,j$ index the coordinates $(x,y,z)=([100],[010],[001])$ in crystallographic notation. Our calculations also found that the components of the $\delta_{hf}$ hyperfine-stress interaction are on the order of $0.1$ $\mathrm{kHz}/\mathrm{kbar}$. Notably, this is four orders of magnitude smaller than the spin-spin dependence on strain (which is on the order of $1$ $\mathrm{MHz}/\mathrm{kbar}$).

To compare with experiment results, we estimate the hyperfine structure under purely hydrostatic pressures. The dependence of $\Delta_{hf}$ with hydrostatic pressure is calculated as follows:

\begin{align}
\frac{\partial\Delta_{hf}}{\partial P} &= \left(\frac{\partial\Delta_{hf}}{\partial \sigma_{xx}} + \frac{\partial\Delta_{hf}}{\partial \sigma_{yy}}+\frac{\partial\Delta_{hf}}{\partial \sigma_{zz}}\right)/3\nonumber\\
&= 0.1186/3 \approx 0.03953 \; \mathrm{MHz}/\mathrm{kbar}.
\end{align}

In the discussion below, the theoretical ambient values are always taken as $\Delta_{hf}(0) = 127.604$ MHz and $\delta_{hf}(0) = 2876.86$ MHz. The dependence of $\delta_{hf}$ on hydrostatic pressure is approximated as the spin-stress interaction $\partial D/\partial P = 1.49 \; \mathrm{MHz}/\mathrm{kbar}$ \cite{Ho2020}. Combining these results, the hyperfine energy levels are approximated by the following expressions:

\begin{align}
\Delta_{hf}(P) &\approx \Delta_{hf}(0) + \frac{\partial \Delta_{hf}}{\partial P} P\nonumber\\
&= 127.604 + 0.03953 P \; \mathrm{MHz},\\
\delta_{hf}(P) &\approx \delta_{hf}(0) + \frac{\partial D}{\partial P} P\nonumber\\
&= 2876.86 + 1.49 P \; \mathrm{MHz}.
\end{align}

\section{Experiments}
\begin{figure}
\includegraphics[width=8.5cm]{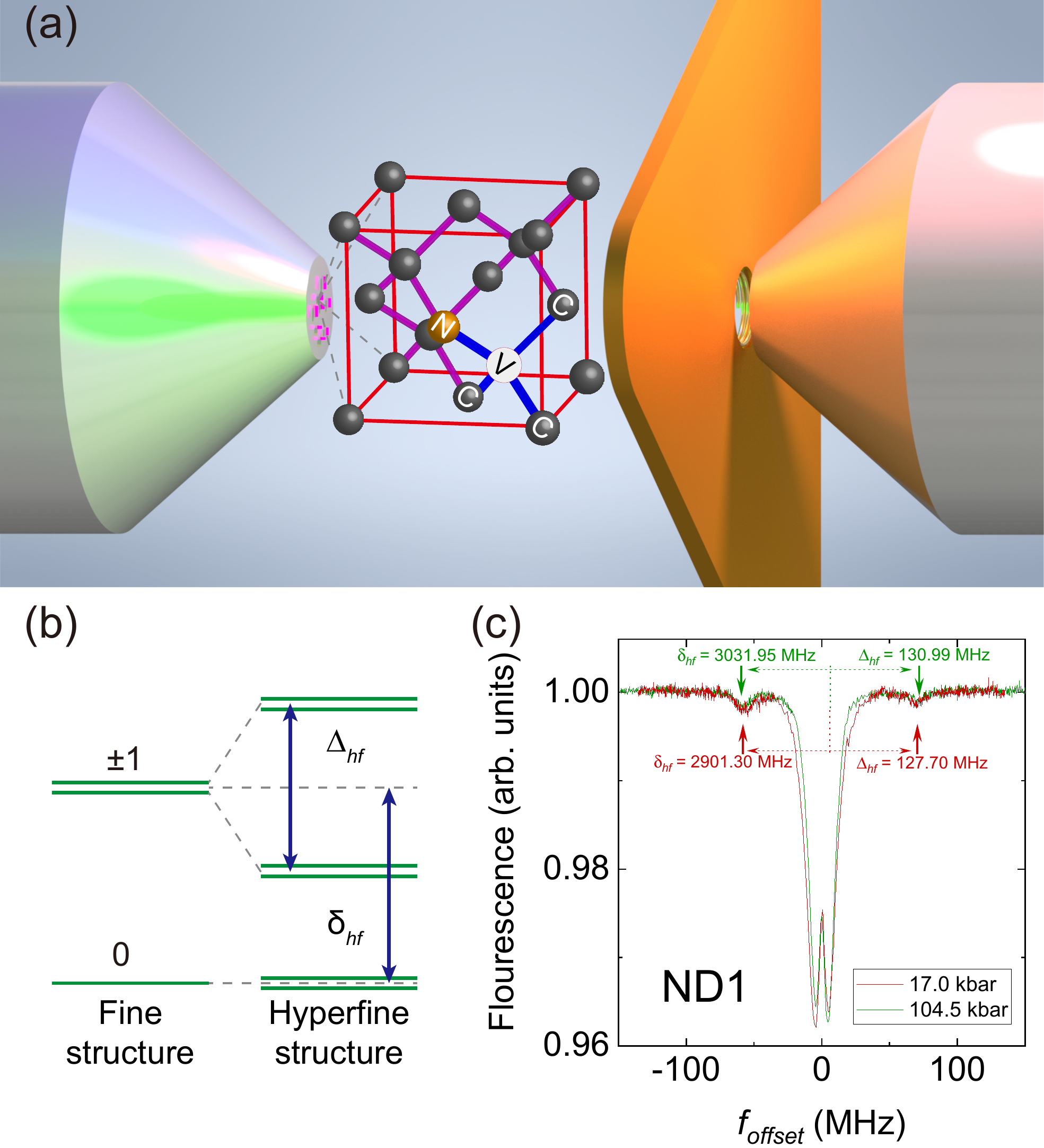}
\caption{(a) An illustration of our DAC. 1-\textmu m NDs were dropcasted onto one of the anvil culets. 4:1 methanol:ethanol mixture was used as the pressure medium that was confined by a metallic gasket, while two aligned diamond anvils were used for generating pressure. The central hole of the gasket had a diameter of 300 \textmu m. A 150-\textmu m-diameter omega-shaped gold micro-structure was used as an MW antenna for ODMR measurements. (b) The energy levels of the hyperfine structures. There are two \ce{^13C} ODMR parameters $\Delta_{hf}$ and $\delta_{hf}$ which correspond to the separation and center of the two hyperfine resonances, respectively. (c) The ODMR spectra of the same ND at 17.0 kbar and 104.5 kbar. The \textit{f\textsubscript{offest}} is defined as the difference from the center frequency $D$. The pressure-induced changes in hyperfine resonances in the ODMR spectrum upon pressure change could be directly observed. It is obvious that the \ce{^13C} hyperfine resonances have relatively broad linewidths and low contrasts compared to the center resonances. The pressure is determined by the center frequency of the corresponding ODMR spectrum \cite{Ho2020}.}
\label{fig1}
\end{figure}

The illustration of our high-pressure device is shown in \cref{fig1}(a), where a diamond anvil cell (DAC) is harnessed to pressurize some 1-\textmu m diamond particles (NDs) drop-casted on one of the anvil culets. To measure spin resonance at high pressures, a 150 \textmu m-diameter omega-shaped gold micro-structure was fabricated on one of the anvils to have uniform and reliable MW transmission \cite{Xie2021}. To ensure an excellent hydrostatic condition up to $\sim$ 100 kbar at room temperature, 4:1 methanol:ethanol mixture was used as the pressure medium \cite{Piermarini1973, Angel2007, Tateiwa2009, Klotz2009}. The pressure is calibrated by individual NDs using $dD/dP = 1.49$ MHz/kbar \cite{Ho2020}. Details are in Ref. \cite{SI}.

\section{Results}
\cref{fig1}(b) shows the energy levels of the \ce{^13C} hyperfine structure. From the ODMR spectrum of NV centers, two parameters could be read out: $\Delta_{hf}$ and $\delta_{hf}$ which correspond to the separation and center of the two \ce{^13C} ODMR resonances, respectively. An example of ODMR spectrum of a 1-\textmu m ND at ambient conditions is shown in Ref. \cite{SI}, and the corresponding hyperfine parameters are $\Delta_{hf}(0) = 127.62 \pm 0.54$ MHz and $\delta_{hf}(0) = 2876.95 \pm 0.38$ MHz. These values are in excellent agreement to the previous studies \cite{Simanovskaia2013, Barson2019, Felton2009, Nizovtsev2010, Mizuochi2009, Baranov2011, Gali2008}. The small discrepancy may be due to internal strain and different charge environments in the \ce{^13C} sample. Therefore, we perform averages over NDs in our data analysis to have statistical results. On the other hand, as the contrast of \ce{^13C} ODMR resonances is relatively weak, peak fitting may contribute a small offset to the result \cite{SI}. The ODMR spectra at 17.0 kbar and 104.5 kbar of the same ND are shown in \cref{fig1}(c). The changes in hyperfine resonances in the ODMR spectrum upon pressure change could be directly observed without any peak-fit processing. To retain high spectral resolution while minimize measurement time, we implemented a non-uniform ODMR measurement to capture the changes. We use a smaller MW frequency step in the hyperfine resonance regions while a larger MW frequency step in the central region. The pressure is determined by the center frequency $D$ of the corresponding ODMR spectrum \cite{Ho2020}.

\begin{figure}
\includegraphics[width=8.5cm]{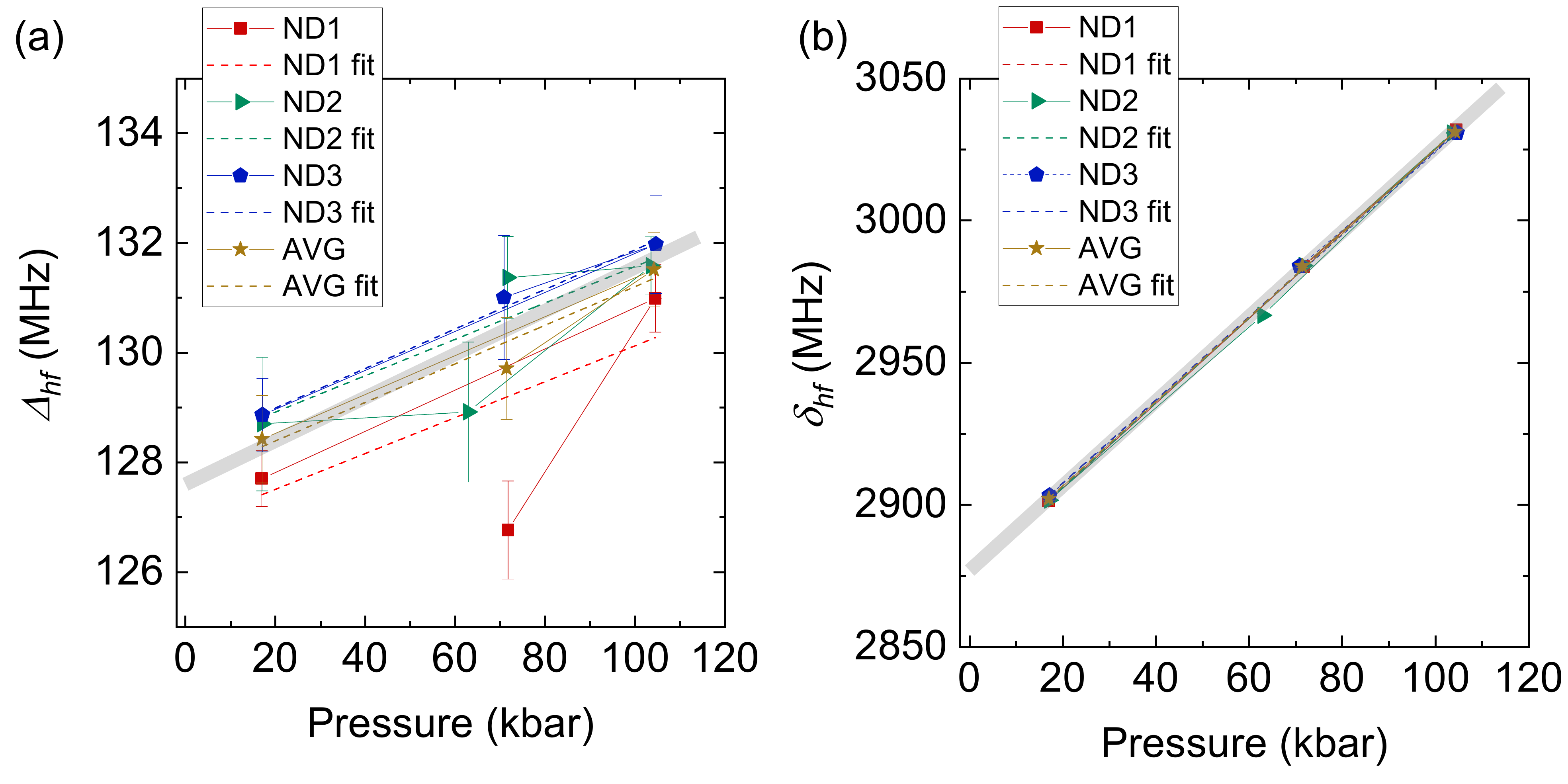}
\caption{Plots of \ce{^{13}C} ODMR parameters (a) $\Delta_{hf}(P)$ and (b) $\delta_{hf}(P)$ against pressure. Three NDs are tracked. The joining lines signify the data measurement sequence. The changes in $\Delta_{hf}$ and $\delta_{hf}$ with pressure are pronounced. The linear fits are the dash lines. The light grey line in the background is the theory result.}
\label{fig2}
\end{figure}

The experimental data of $\Delta_{hf}(P)$ and $\delta_{hf}(P)$ at different pressures are plotted in \cref{fig2}. Three NDs are tracked. We have tried three different ways to process the data of $\Delta_{hf}(P)$ and $\delta_{hf}(P)$: (i) perform a linear fit for each of the three NDs; (ii) take the average of the three then perform a linear fit (AVG); and (iii) take the average after performing individual linear fits (AVG2). The changes in $\Delta_{hf}$ and $\delta_{hf}$ upon pressure change are pronounced. When pressure increases, the \ce{^{13}C} ODMR parameters $\Delta_{hf}$ and $\delta_{hf}$ gradually increase. For $\Delta_{hf}$, the data are slightly scattered among different NDs but the same increasing behavior is noted. The discrepancy is from the local charge environment and lattice strain, as well as tiny strain perturbation projected to the NV local frame. For $\delta_{hf}$, the data points fall nicely on the same linear line. Note that the pressure dependence of $\delta_{hf}(P)$ is dominated by the spin-spin dependence ($dD/dP$), hence the discrepancy in $\delta_{hf}(P)$ is much smaller compared to $\Delta_{hf}(P)$. These results are summarized in \cref{values} and they are in great agreement with the theoretical calculation. \cref{fig2} is strong evidence showing that the hyperfine parameters $A_{\|}$ and $A_{\perp}$ change under pressure. Additional data are shown in the Ref. \cite{SI}.

\begin{table}[h!]
\centering
\begin{tabular}{|c|c|c|c|c|}
\hline
Label & $\Delta_{hf}(0)$ & $d\Delta_{hf}/dP$ & $\delta_{hf}(0)$ & d$\delta_{hf}$/d$P$
\\ \hline
ND1 & 126.85 & 0.033 & 2876.17 & 1.492
\\ \hline
ND2 & 128.25 & 0.033 & 2874.96 & 1.500
\\ \hline
ND3 & 128.27 & 0.036 & 2878.41 & 1.463
\\ \hline
AVG & 127.69 & 0.035 & 2877.05 & 1.482
\\ \hline
AVG2 & 127.79 & 0.034 & 2876.51 & 1.485
\\ \hline
Theory & 127.604 & 0.03953 & 2876.86 & 1.49
\\ \hline
\end{tabular}
\caption{Data summary of the \ce{^{13}C} ODMR parameters $\Delta_{hf}(P)$ and $\delta_{hf}(P)$. All values are in MHz unit. Our data are in great agreement with the theory.}
\label{values}
\end{table}

\begin{figure}
\includegraphics[width=8.5cm]{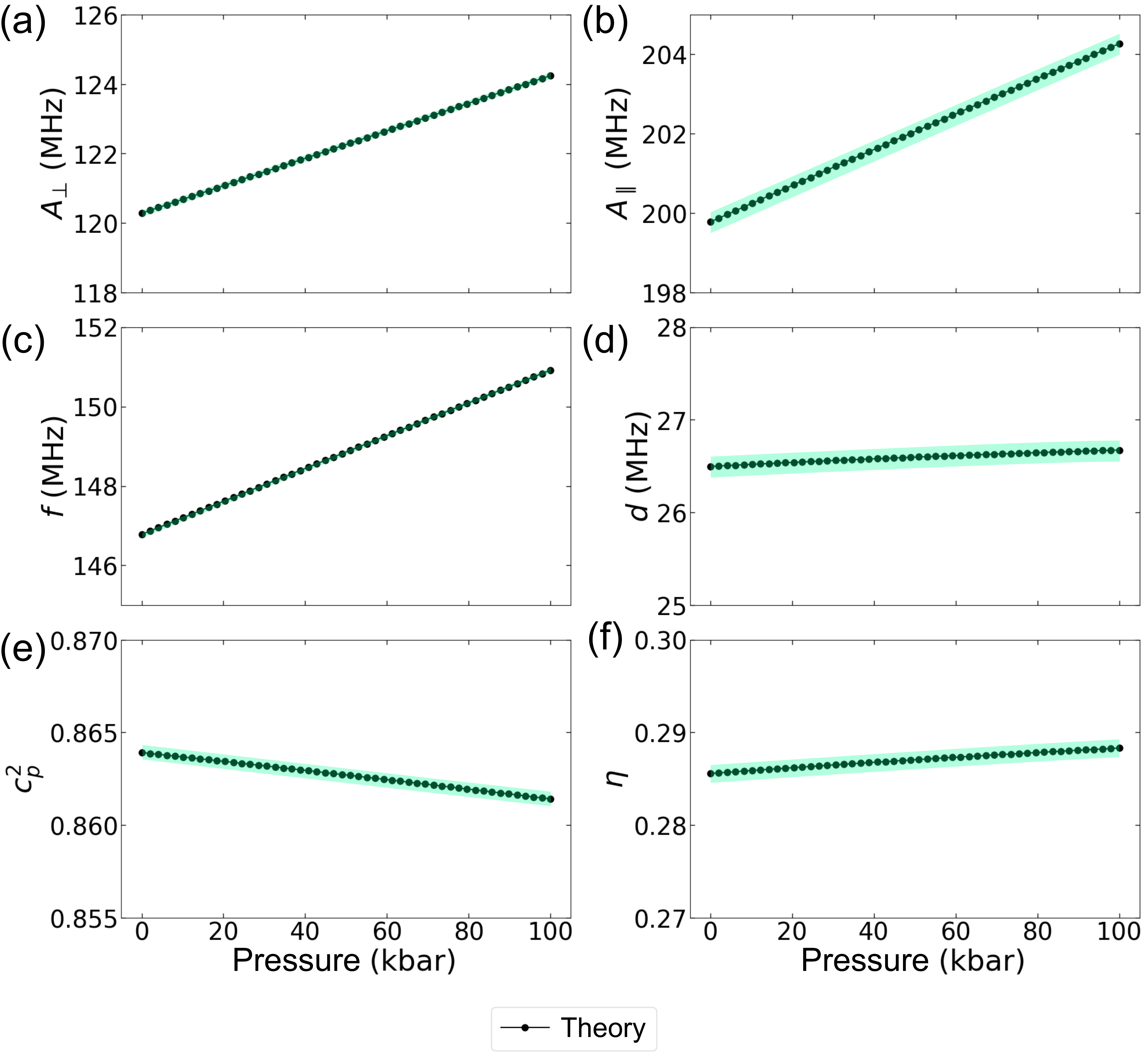}
\caption{Derived results of (a) hyperfine constant $A_\perp$, (b) hyperfine constant $A_{\|}$, (c) Fermi contact term $f$, (d) dipole term $d$, (e) hybridization of the $p$ orbitals $\abs{c_{p}}^2$, and (f) electron spin density $\eta$. The black markers are derived purely from theoretical values. The electron density $\eta$ increases while the $p$-orbital contribution $\abs{c_{p}}^2$ decreases, both slightly. The green error bound is from the maximum uncertainty of the theoretical resolution of 0.1 kHz/kbar. This theoretical resolution is the smallest energy difference resolvable in our calculations. It mainly affects the terms $d\delta_{hf}/dP$ since the neglected hyperfine-stress interaction in $\delta_{hf}$ is of the same order. Furthermore, the term $d\delta_{hf}/dP$ is shown to be the most sensitive input parameter that impacts the derived results \cite{SI}. Note that the theoretical resolution of 0.1 kHz/kbar is much better than the test case of, yet still merely, 0.1\% difference (around 1 kHz/kbar).}
\label{fig3}
\end{figure}

\begin{figure}
\includegraphics[width=8.5cm]{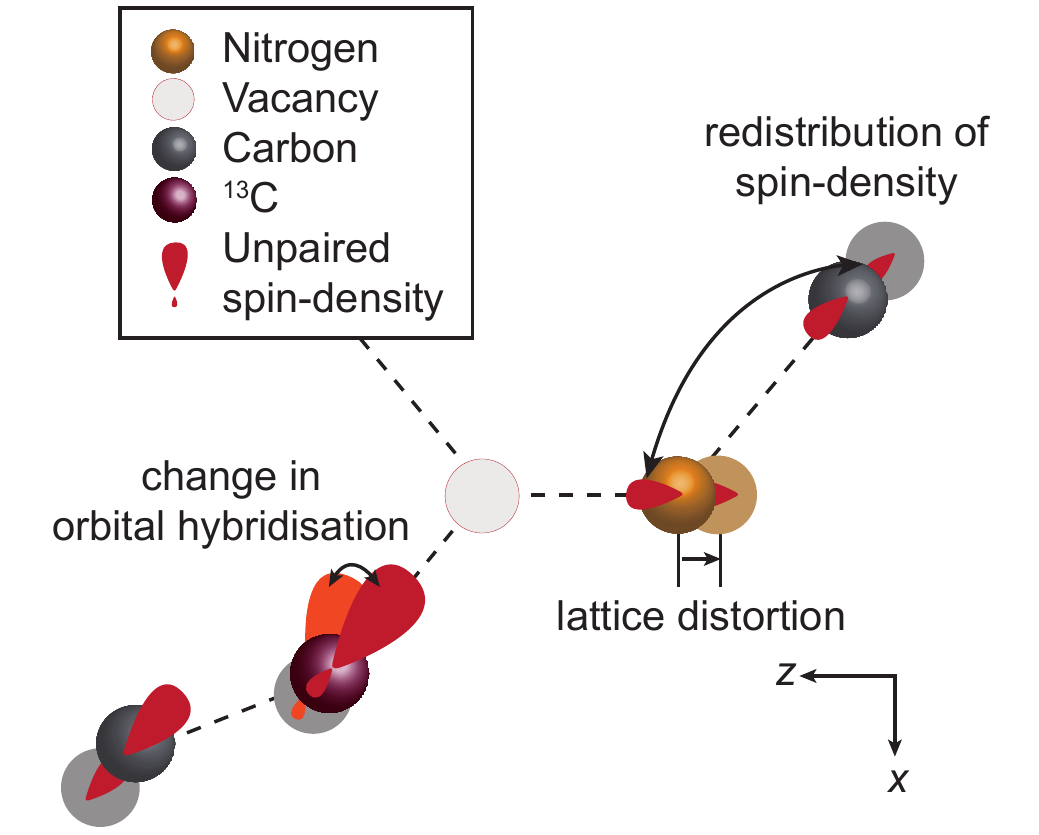}
\caption{Schematic of lattice distortion and the consequent redistribution of spin density between atoms and rehybridization of the atomic orbitals. The arrows indicate the displacements of atoms for increasing pressure. The axes $(X, Z)$ correspond to the crystallographic coordinate system.}
\label{fig4}
\end{figure}

Our results reveal a prominent change in the hyperfine parameters $A_{\|}$ and $A_{\perp}$, thus suggesting a possible variation in the NV electron spin wavefunction upon applying pressure. To quantify the changes in the NV electron spin density and orbital hybridization, the \ce{^{13}C} ODMR parameters $\Delta_{hf}$ and $\delta_{hf}$ were first converted to hyperfine parameters $A_{\|}$ and $A_{\perp}$, then further to the Fermi contact \textit{f} and dipole term \textit{d}, and finally, to the orbital hybridization $\abs{c_{s}}^{2}/\abs{c_{p}}^{2}$ and electron spin density $\eta$. The derived results are shown in \cref{fig3}, which indicates that under applied hydrostatic pressure the electron density increases while the defect orbitals become more $s$-like and less $p$-like. Applying a qualitative model of the defect, we expect the orbitals to become more localized to the atoms around the vacancy. This is because pressure reduces the size of lattice unit cell, resulting in a deepening of the potential well of the vacancy and hence an increased attraction of electron density. We also expect that the NV center self-distorts away from a tetrahedral configuration under pressure. As a result, the bonds between the nearest neighbor and next-to-nearest carbon atoms become more like $sp^2$ bonds and less like $sp^3$ bonds. This is consistent with the decrease in the $p$-orbital contribution to the bonds, $\abs{c_{p}}^{2}$. \cref{fig4} illustrates the lattice distortion under pressure which gives rise to changes in orbital hybridization of the unpaired NV spin and electron spin density at the location of the nuclear spin.

\section{Discussions}
The nonlinearity nature of solving the Hamiltonian backwards from $\Delta_{hf}$ and $\delta_{hf}$ to $A_{\|}$ and $A_{\perp}$ leads to sensitive dependence on certain parameters. Although our measurement data match the theoretical values pretty well, small differences are severely amplified by the non-linear equations. We have tested all four \ce{^{13}C} ODMR parameters (slopes and intercepts of $\Delta_{hf}(P)$ and $\delta_{hf}(P)$) and studied how sensitive are the six derived quantities in \cref{fig3} to them. The analysis is summarized in Ref. \cite{SI}. Surprisingly, changing $d\Delta_{hf}/dP$ does not alter the calculations by much, even changing $\Delta_{hf}(0)$ seems to only offset the calculations by a little amount. However, $d\delta_{hf}/dP$ affects the calculations significantly. Merely changing $d\delta_{hf}/dP$ by 0.1\% (around 1 kHz/kbar) can significantly modify the result, even giving different signs of the slopes of the electron spin density $\eta(P)$ and $p$-orbital hybridization $\abs{c_{p}(P)}^2$. The components of the $\delta_{hf}$ hyperfine-stress interaction are on the
order of 0.1 kHz/kbar as stated in the theory section, and hence a seemingly small change of 1 kHz/kbar in $d\delta_{hf}/dP$ actually matters. The change in $\delta_{hf}(0)$ also offsets the values significantly, but the trends are similar.

In principle, the experimental resolution can be improved by ENDOR Raman-heterodyne \cite{Manson1990} or pulsed NMR measurements. However, the low abundance of \ce{^13C} limits the sensitivity. By using \ce{^13C} enriched samples, one can reach higher accuracy and the problem of low \ce{^13C} concentration may not be a limitation in other defect systems. Furthermore, we are using a commercial ND product, while further work will require custom specifications to enhance the resolution.

In future work, HSE calculations can be conducted to further reveal the localization of electronic states and hyperfine parameters \cite{Szasz2013Hyperfine, Swift2020First}. The accuracy of PBE NV hyperfine is due to a fortuitous cancellation of two main effects. First, neglecting the spin-polarization of the core electrons results in a systematic overestimation of the hyperfine interaction. Second, the spin-density calculated with PBE is less localized resulting in a systematic underestimation of the hyperfine interaction. Previous work has identified systematic overestimation and underestimation cancel to fortuitously give the near-exact agreement seen in PBE. While this fortuitous cancellation may persist under applied strain, the discrepancy will be bounded by about 20\%.

Note that the thermal and pressure effects are similar but not identical. While pressure and temperature can change lattice constant, temperature also changes the distribution of phonon modes, hence changing the optical and spin properties. From this perspective, pressure tuning is a cleaner approach to have an isolated effect on the parameters for systematic studies. Actually, by combing both approaches, i.e. scanning temperature in this pressure method, one may decouple the thermal/phonon effect. Moreover, a pressure cell is a small portable device and is more robust than a cooler or heater.  Meanwhile, the pressure tuning method is orders of magnitude stronger than thermal one, providing a much wider scanning range of lattice changes.

In summary, we have revealed the pressure dependence of \ce{^13C} hyperfine structure. In particular, we focus on the hyperfine interaction between the NV electron spin and the \ce{^13C} nuclear spin, which leads to the detection of the NV center electron spin density and orbitals. Under applied hydrostatic pressure, the electron spin density $\eta$ increases while the defect orbitals become more like $sp^2$ bonds and less like $sp^3$ bonds. These can be explained by the reduction of the cell size and the self-distortion of the NV center. By using pressure tuning and \ce{^{13}C} nuclear spins as a probe, we demonstrate that one can investigate the wavefunctions of deep defects in semiconductors. This method can be adopted to study similar defect systems with hyperfine structure. Hence, we can adopt this process to understand and optimize the quantum sensing and computing applications of a wide array of deep defects in semiconductors.

\begin{acknowledgments}
We thank K. P. Ao for the technical support. We thank P. T. Fong for the discussion. K.O.H acknowledges financial support from the Hong Kong PhD Fellowship Scheme. S.K.G. acknowledges financial support from Hong Kong RGC (GRF/14300418, GRF/14301020, and A-CUHK402/19). S.Y. acknowledges financial support from Hong Kong RGC (GRF/14304419).
\end{acknowledgments}

\bibliography{hyperfine_pressure}

\begin{thebibliography}{57}%
\makeatletter
\providecommand \@ifxundefined [1]{%
 \@ifx{#1\undefined}
}%
\providecommand \@ifnum [1]{%
 \ifnum #1\expandafter \@firstoftwo
 \else \expandafter \@secondoftwo
 \fi
}%
\providecommand \@ifx [1]{%
 \ifx #1\expandafter \@firstoftwo
 \else \expandafter \@secondoftwo
 \fi
}%
\providecommand \natexlab [1]{#1}%
\providecommand \enquote  [1]{``#1''}%
\providecommand \bibnamefont  [1]{#1}%
\providecommand \bibfnamefont [1]{#1}%
\providecommand \citenamefont [1]{#1}%
\providecommand \href@noop [0]{\@secondoftwo}%
\providecommand \href [0]{\begingroup \@sanitize@url \@href}%
\providecommand \@href[1]{\@@startlink{#1}\@@href}%
\providecommand \@@href[1]{\endgroup#1\@@endlink}%
\providecommand \@sanitize@url [0]{\catcode `\\12\catcode `\$12\catcode
  `\&12\catcode `\#12\catcode `\^12\catcode `\_12\catcode `\%12\relax}%
\providecommand \@@startlink[1]{}%
\providecommand \@@endlink[0]{}%
\providecommand \url  [0]{\begingroup\@sanitize@url \@url }%
\providecommand \@url [1]{\endgroup\@href {#1}{\urlprefix }}%
\providecommand \urlprefix  [0]{URL }%
\providecommand \Eprint [0]{\href }%
\providecommand \doibase [0]{https://doi.org/}%
\providecommand \selectlanguage [0]{\@gobble}%
\providecommand \bibinfo  [0]{\@secondoftwo}%
\providecommand \bibfield  [0]{\@secondoftwo}%
\providecommand \translation [1]{[#1]}%
\providecommand \BibitemOpen [0]{}%
\providecommand \bibitemStop [0]{}%
\providecommand \bibitemNoStop [0]{.\EOS\space}%
\providecommand \EOS [0]{\spacefactor3000\relax}%
\providecommand \BibitemShut  [1]{\csname bibitem#1\endcsname}%
\let\auto@bib@innerbib\@empty
\bibitem [{\citenamefont {Barson}\ \emph {et~al.}(2017)\citenamefont {Barson},
  \citenamefont {Peddibhotla}, \citenamefont {Ovartchaiyapong}, \citenamefont
  {Ganesan}, \citenamefont {Taylor}, \citenamefont {Gebert}, \citenamefont
  {Mielens}, \citenamefont {Koslowski}, \citenamefont {Simpson}, \citenamefont
  {McGuinness}, \citenamefont {McCallum}, \citenamefont {Prawer}, \citenamefont
  {Onoda}, \citenamefont {Ohshima}, \citenamefont {Bleszynski~Jayich},
  \citenamefont {Jelezko}, \citenamefont {Manson},\ and\ \citenamefont
  {Doherty}}]{Barson2017}%
  \BibitemOpen
  \bibfield  {author} {\bibinfo {author} {\bibfnamefont {M.~S.~J.}\
  \bibnamefont {Barson}}, \bibinfo {author} {\bibfnamefont {P.}~\bibnamefont
  {Peddibhotla}}, \bibinfo {author} {\bibfnamefont {P.}~\bibnamefont
  {Ovartchaiyapong}}, \bibinfo {author} {\bibfnamefont {K.}~\bibnamefont
  {Ganesan}}, \bibinfo {author} {\bibfnamefont {R.~L.}\ \bibnamefont {Taylor}},
  \bibinfo {author} {\bibfnamefont {M.}~\bibnamefont {Gebert}}, \bibinfo
  {author} {\bibfnamefont {Z.}~\bibnamefont {Mielens}}, \bibinfo {author}
  {\bibfnamefont {B.}~\bibnamefont {Koslowski}}, \bibinfo {author}
  {\bibfnamefont {D.~A.}\ \bibnamefont {Simpson}}, \bibinfo {author}
  {\bibfnamefont {L.~P.}\ \bibnamefont {McGuinness}}, \bibinfo {author}
  {\bibfnamefont {J.}~\bibnamefont {McCallum}}, \bibinfo {author}
  {\bibfnamefont {S.}~\bibnamefont {Prawer}}, \bibinfo {author} {\bibfnamefont
  {S.}~\bibnamefont {Onoda}}, \bibinfo {author} {\bibfnamefont
  {T.}~\bibnamefont {Ohshima}}, \bibinfo {author} {\bibfnamefont {A.~C.}\
  \bibnamefont {Bleszynski~Jayich}}, \bibinfo {author} {\bibfnamefont
  {F.}~\bibnamefont {Jelezko}}, \bibinfo {author} {\bibfnamefont {N.~B.}\
  \bibnamefont {Manson}},\ and\ \bibinfo {author} {\bibfnamefont {M.~W.}\
  \bibnamefont {Doherty}},\ }\bibfield  {title} {\bibinfo {title}
  {Nanomechanical sensing using spins in diamond},\ }\href
  {https://doi.org/10.1021/acs.nanolett.6b04544} {\bibfield  {journal}
  {\bibinfo  {journal} {Nano Letters}\ }\textbf {\bibinfo {volume} {17}},\
  \bibinfo {pages} {1496} (\bibinfo {year} {2017})},\ \bibinfo {note} {pMID:
  28146361},\ \Eprint
  {https://arxiv.org/abs/https://doi.org/10.1021/acs.nanolett.6b04544}
  {https://doi.org/10.1021/acs.nanolett.6b04544} \BibitemShut {NoStop}%
\bibitem [{\citenamefont {Doherty}\ \emph
  {et~al.}(2014{\natexlab{a}})\citenamefont {Doherty}, \citenamefont
  {Struzhkin}, \citenamefont {Simpson}, \citenamefont {McGuinness},
  \citenamefont {Meng}, \citenamefont {Stacey}, \citenamefont {Karle},
  \citenamefont {Hemley}, \citenamefont {Manson}, \citenamefont {Hollenberg},\
  and\ \citenamefont {Prawer}}]{Doherty2014P}%
  \BibitemOpen
  \bibfield  {author} {\bibinfo {author} {\bibfnamefont {M.~W.}\ \bibnamefont
  {Doherty}}, \bibinfo {author} {\bibfnamefont {V.~V.}\ \bibnamefont
  {Struzhkin}}, \bibinfo {author} {\bibfnamefont {D.~A.}\ \bibnamefont
  {Simpson}}, \bibinfo {author} {\bibfnamefont {L.~P.}\ \bibnamefont
  {McGuinness}}, \bibinfo {author} {\bibfnamefont {Y.}~\bibnamefont {Meng}},
  \bibinfo {author} {\bibfnamefont {A.}~\bibnamefont {Stacey}}, \bibinfo
  {author} {\bibfnamefont {T.~J.}\ \bibnamefont {Karle}}, \bibinfo {author}
  {\bibfnamefont {R.~J.}\ \bibnamefont {Hemley}}, \bibinfo {author}
  {\bibfnamefont {N.~B.}\ \bibnamefont {Manson}}, \bibinfo {author}
  {\bibfnamefont {L.~C.~L.}\ \bibnamefont {Hollenberg}},\ and\ \bibinfo
  {author} {\bibfnamefont {S.}~\bibnamefont {Prawer}},\ }\bibfield  {title}
  {\bibinfo {title} {Electronic properties and metrology applications of the
  diamond ${\mathrm{nv}}^{\ensuremath{-}}$ center under pressure},\ }\href
  {https://doi.org/10.1103/PhysRevLett.112.047601} {\bibfield  {journal}
  {\bibinfo  {journal} {Phys. Rev. Lett.}\ }\textbf {\bibinfo {volume} {112}},\
  \bibinfo {pages} {047601} (\bibinfo {year} {2014}{\natexlab{a}})}\BibitemShut
  {NoStop}%
\bibitem [{\citenamefont {Barson}\ \emph {et~al.}(2019)\citenamefont {Barson},
  \citenamefont {Reddy}, \citenamefont {Yang}, \citenamefont {Manson},
  \citenamefont {Wrachtrup},\ and\ \citenamefont {Doherty}}]{Barson2019}%
  \BibitemOpen
  \bibfield  {author} {\bibinfo {author} {\bibfnamefont {M.~S.~J.}\
  \bibnamefont {Barson}}, \bibinfo {author} {\bibfnamefont {P.}~\bibnamefont
  {Reddy}}, \bibinfo {author} {\bibfnamefont {S.}~\bibnamefont {Yang}},
  \bibinfo {author} {\bibfnamefont {N.~B.}\ \bibnamefont {Manson}}, \bibinfo
  {author} {\bibfnamefont {J.}~\bibnamefont {Wrachtrup}},\ and\ \bibinfo
  {author} {\bibfnamefont {M.~W.}\ \bibnamefont {Doherty}},\ }\bibfield
  {title} {\bibinfo {title} {Temperature dependence of the $^{13}\mathrm{C}$
  hyperfine structure of the negatively charged nitrogen-vacancy center in
  diamond},\ }\href {https://doi.org/10.1103/PhysRevB.99.094101} {\bibfield
  {journal} {\bibinfo  {journal} {Phys. Rev. B}\ }\textbf {\bibinfo {volume}
  {99}},\ \bibinfo {pages} {094101} (\bibinfo {year} {2019})}\BibitemShut
  {NoStop}%
\bibitem [{\citenamefont {Soshenko}\ \emph {et~al.}(2020)\citenamefont
  {Soshenko}, \citenamefont {Vorobyov}, \citenamefont {Bolshedvorskii},
  \citenamefont {Rubinas}, \citenamefont {Cojocaru}, \citenamefont {Kudlatsky},
  \citenamefont {Zeleneev}, \citenamefont {Sorokin}, \citenamefont
  {Smolyaninov},\ and\ \citenamefont {Akimov}}]{soshenko2020Temperature}%
  \BibitemOpen
  \bibfield  {author} {\bibinfo {author} {\bibfnamefont {V.~V.}\ \bibnamefont
  {Soshenko}}, \bibinfo {author} {\bibfnamefont {V.~V.}\ \bibnamefont
  {Vorobyov}}, \bibinfo {author} {\bibfnamefont {S.~V.}\ \bibnamefont
  {Bolshedvorskii}}, \bibinfo {author} {\bibfnamefont {O.}~\bibnamefont
  {Rubinas}}, \bibinfo {author} {\bibfnamefont {I.}~\bibnamefont {Cojocaru}},
  \bibinfo {author} {\bibfnamefont {B.}~\bibnamefont {Kudlatsky}}, \bibinfo
  {author} {\bibfnamefont {A.~I.}\ \bibnamefont {Zeleneev}}, \bibinfo {author}
  {\bibfnamefont {V.~N.}\ \bibnamefont {Sorokin}}, \bibinfo {author}
  {\bibfnamefont {A.~N.}\ \bibnamefont {Smolyaninov}},\ and\ \bibinfo {author}
  {\bibfnamefont {A.~V.}\ \bibnamefont {Akimov}},\ }\bibfield  {title}
  {\bibinfo {title} {Temperature drift rate for nuclear terms of the nv-center
  ground-state hamiltonian},\ }\href
  {https://doi.org/10.1103/PhysRevB.102.125133} {\bibfield  {journal} {\bibinfo
   {journal} {Phys. Rev. B}\ }\textbf {\bibinfo {volume} {102}},\ \bibinfo
  {pages} {125133} (\bibinfo {year} {2020})}\BibitemShut {NoStop}%
\bibitem [{\citenamefont {Jarmola}\ \emph {et~al.}(2020)\citenamefont
  {Jarmola}, \citenamefont {Fescenko}, \citenamefont {Acosta}, \citenamefont
  {Doherty}, \citenamefont {Fatemi}, \citenamefont {Ivanov}, \citenamefont
  {Budker},\ and\ \citenamefont {Malinovsky}}]{Jarmola2020Robust}%
  \BibitemOpen
  \bibfield  {author} {\bibinfo {author} {\bibfnamefont {A.}~\bibnamefont
  {Jarmola}}, \bibinfo {author} {\bibfnamefont {I.}~\bibnamefont {Fescenko}},
  \bibinfo {author} {\bibfnamefont {V.~M.}\ \bibnamefont {Acosta}}, \bibinfo
  {author} {\bibfnamefont {M.~W.}\ \bibnamefont {Doherty}}, \bibinfo {author}
  {\bibfnamefont {F.~K.}\ \bibnamefont {Fatemi}}, \bibinfo {author}
  {\bibfnamefont {T.}~\bibnamefont {Ivanov}}, \bibinfo {author} {\bibfnamefont
  {D.}~\bibnamefont {Budker}},\ and\ \bibinfo {author} {\bibfnamefont {V.~S.}\
  \bibnamefont {Malinovsky}},\ }\bibfield  {title} {\bibinfo {title} {Robust
  optical readout and characterization of nuclear spin transitions in
  nitrogen-vacancy ensembles in diamond},\ }\href
  {https://doi.org/10.1103/PhysRevResearch.2.023094} {\bibfield  {journal}
  {\bibinfo  {journal} {Phys. Rev. Research}\ }\textbf {\bibinfo {volume}
  {2}},\ \bibinfo {pages} {023094} (\bibinfo {year} {2020})}\BibitemShut
  {NoStop}%
\bibitem [{\citenamefont {Wang}\ \emph {et~al.}(2022)\citenamefont {Wang},
  \citenamefont {Barr}, \citenamefont {Tang}, \citenamefont {Chen},
  \citenamefont {Li}, \citenamefont {Xu}, \citenamefont {Li},\ and\
  \citenamefont {Cappellaro}}]{Wang2022Characterizing}%
  \BibitemOpen
  \bibfield  {author} {\bibinfo {author} {\bibfnamefont {G.}~\bibnamefont
  {Wang}}, \bibinfo {author} {\bibfnamefont {A.~R.}\ \bibnamefont {Barr}},
  \bibinfo {author} {\bibfnamefont {H.}~\bibnamefont {Tang}}, \bibinfo {author}
  {\bibfnamefont {M.}~\bibnamefont {Chen}}, \bibinfo {author} {\bibfnamefont
  {C.}~\bibnamefont {Li}}, \bibinfo {author} {\bibfnamefont {H.}~\bibnamefont
  {Xu}}, \bibinfo {author} {\bibfnamefont {J.}~\bibnamefont {Li}},\ and\
  \bibinfo {author} {\bibfnamefont {P.}~\bibnamefont {Cappellaro}},\ }\href
  {https://doi.org/10.48550/ARXIV.2205.02790} {\bibinfo {title} {Characterizing
  temperature and strain variations with qubit ensembles for their robust
  coherence protection}} (\bibinfo {year} {2022})\BibitemShut {NoStop}%
\bibitem [{\citenamefont {Tang}\ \emph {et~al.}(2022)\citenamefont {Tang},
  \citenamefont {Barr}, \citenamefont {Wang}, \citenamefont {Cappellaro},\ and\
  \citenamefont {Li}}]{Tang2022First}%
  \BibitemOpen
  \bibfield  {author} {\bibinfo {author} {\bibfnamefont {H.}~\bibnamefont
  {Tang}}, \bibinfo {author} {\bibfnamefont {A.~R.}\ \bibnamefont {Barr}},
  \bibinfo {author} {\bibfnamefont {G.}~\bibnamefont {Wang}}, \bibinfo {author}
  {\bibfnamefont {P.}~\bibnamefont {Cappellaro}},\ and\ \bibinfo {author}
  {\bibfnamefont {J.}~\bibnamefont {Li}},\ }\href
  {https://doi.org/10.48550/ARXIV.2205.02791} {\bibinfo {title}
  {First-principles calculation of the temperature-dependent transition
  energies in spin defects}} (\bibinfo {year} {2022})\BibitemShut {NoStop}%
\bibitem [{\citenamefont {Yip}\ \emph {et~al.}(2019)\citenamefont {Yip},
  \citenamefont {Ho}, \citenamefont {Yu}, \citenamefont {Chen}, \citenamefont
  {Zhang}, \citenamefont {Kasahara}, \citenamefont {Mizukami}, \citenamefont
  {Shibauchi}, \citenamefont {Matsuda}, \citenamefont {Goh},\ and\
  \citenamefont {Yang}}]{Yip2019}%
  \BibitemOpen
  \bibfield  {author} {\bibinfo {author} {\bibfnamefont {K.~Y.}\ \bibnamefont
  {Yip}}, \bibinfo {author} {\bibfnamefont {K.~O.}\ \bibnamefont {Ho}},
  \bibinfo {author} {\bibfnamefont {K.~Y.}\ \bibnamefont {Yu}}, \bibinfo
  {author} {\bibfnamefont {Y.}~\bibnamefont {Chen}}, \bibinfo {author}
  {\bibfnamefont {W.}~\bibnamefont {Zhang}}, \bibinfo {author} {\bibfnamefont
  {S.}~\bibnamefont {Kasahara}}, \bibinfo {author} {\bibfnamefont
  {Y.}~\bibnamefont {Mizukami}}, \bibinfo {author} {\bibfnamefont
  {T.}~\bibnamefont {Shibauchi}}, \bibinfo {author} {\bibfnamefont
  {Y.}~\bibnamefont {Matsuda}}, \bibinfo {author} {\bibfnamefont {S.~K.}\
  \bibnamefont {Goh}},\ and\ \bibinfo {author} {\bibfnamefont {S.}~\bibnamefont
  {Yang}},\ }\bibfield  {title} {\bibinfo {title} {Measuring magnetic field
  texture in correlated electron systems under extreme conditions},\ }\href
  {https://doi.org/10.1126/science.aaw4278} {\bibfield  {journal} {\bibinfo
  {journal} {Science}\ }\textbf {\bibinfo {volume} {366}},\ \bibinfo {pages}
  {1355} (\bibinfo {year} {2019})},\ \Eprint
  {https://arxiv.org/abs/https://science.sciencemag.org/content/366/6471/1355.full.pdf}
  {https://science.sciencemag.org/content/366/6471/1355.full.pdf} \BibitemShut
  {NoStop}%
\bibitem [{\citenamefont {Ho}\ \emph {et~al.}(2020)\citenamefont {Ho},
  \citenamefont {Leung}, \citenamefont {Jiang}, \citenamefont {Ao},
  \citenamefont {Zhang}, \citenamefont {Yip}, \citenamefont {Pang},
  \citenamefont {Wong}, \citenamefont {Goh},\ and\ \citenamefont
  {Yang}}]{Ho2020}%
  \BibitemOpen
  \bibfield  {author} {\bibinfo {author} {\bibfnamefont {K.~O.}\ \bibnamefont
  {Ho}}, \bibinfo {author} {\bibfnamefont {M.~Y.}\ \bibnamefont {Leung}},
  \bibinfo {author} {\bibfnamefont {Y.}~\bibnamefont {Jiang}}, \bibinfo
  {author} {\bibfnamefont {K.~P.}\ \bibnamefont {Ao}}, \bibinfo {author}
  {\bibfnamefont {W.}~\bibnamefont {Zhang}}, \bibinfo {author} {\bibfnamefont
  {K.~Y.}\ \bibnamefont {Yip}}, \bibinfo {author} {\bibfnamefont {Y.~Y.}\
  \bibnamefont {Pang}}, \bibinfo {author} {\bibfnamefont {K.~C.}\ \bibnamefont
  {Wong}}, \bibinfo {author} {\bibfnamefont {S.~K.}\ \bibnamefont {Goh}},\ and\
  \bibinfo {author} {\bibfnamefont {S.}~\bibnamefont {Yang}},\ }\bibfield
  {title} {\bibinfo {title} {Probing local pressure environment in anvil cells
  with nitrogen-vacancy (n-${V}^{\ensuremath{-}}$) centers in diamond},\ }\href
  {https://doi.org/10.1103/PhysRevApplied.13.024041} {\bibfield  {journal}
  {\bibinfo  {journal} {Phys. Rev. Applied}\ }\textbf {\bibinfo {volume}
  {13}},\ \bibinfo {pages} {024041} (\bibinfo {year} {2020})}\BibitemShut
  {NoStop}%
\bibitem [{\citenamefont {Lesik}\ \emph {et~al.}(2019)\citenamefont {Lesik},
  \citenamefont {Plisson}, \citenamefont {Toraille}, \citenamefont {Renaud},
  \citenamefont {Occelli}, \citenamefont {Schmidt}, \citenamefont {Salord},
  \citenamefont {Delobbe}, \citenamefont {Debuisschert}, \citenamefont
  {Rondin}, \citenamefont {Loubeyre},\ and\ \citenamefont {Roch}}]{Lesik2019}%
  \BibitemOpen
  \bibfield  {author} {\bibinfo {author} {\bibfnamefont {M.}~\bibnamefont
  {Lesik}}, \bibinfo {author} {\bibfnamefont {T.}~\bibnamefont {Plisson}},
  \bibinfo {author} {\bibfnamefont {L.}~\bibnamefont {Toraille}}, \bibinfo
  {author} {\bibfnamefont {J.}~\bibnamefont {Renaud}}, \bibinfo {author}
  {\bibfnamefont {F.}~\bibnamefont {Occelli}}, \bibinfo {author} {\bibfnamefont
  {M.}~\bibnamefont {Schmidt}}, \bibinfo {author} {\bibfnamefont
  {O.}~\bibnamefont {Salord}}, \bibinfo {author} {\bibfnamefont
  {A.}~\bibnamefont {Delobbe}}, \bibinfo {author} {\bibfnamefont
  {T.}~\bibnamefont {Debuisschert}}, \bibinfo {author} {\bibfnamefont
  {L.}~\bibnamefont {Rondin}}, \bibinfo {author} {\bibfnamefont
  {P.}~\bibnamefont {Loubeyre}},\ and\ \bibinfo {author} {\bibfnamefont
  {J.-F.}\ \bibnamefont {Roch}},\ }\bibfield  {title} {\bibinfo {title}
  {Magnetic measurements on micrometer-sized samples under high pressure using
  designed nv centers},\ }\href {https://doi.org/10.1126/science.aaw4329}
  {\bibfield  {journal} {\bibinfo  {journal} {Science}\ }\textbf {\bibinfo
  {volume} {366}},\ \bibinfo {pages} {1359} (\bibinfo {year} {2019})},\ \Eprint
  {https://arxiv.org/abs/https://science.sciencemag.org/content/366/6471/1359.full.pdf}
  {https://science.sciencemag.org/content/366/6471/1359.full.pdf} \BibitemShut
  {NoStop}%
\bibitem [{\citenamefont {Hsieh}\ \emph {et~al.}(2019)\citenamefont {Hsieh},
  \citenamefont {Bhattacharyya}, \citenamefont {Zu}, \citenamefont {Mittiga},
  \citenamefont {Smart}, \citenamefont {Machado}, \citenamefont {Kobrin},
  \citenamefont {H{\"o}hn}, \citenamefont {Rui}, \citenamefont {Kamrani},
  \citenamefont {Chatterjee}, \citenamefont {Choi}, \citenamefont {Zaletel},
  \citenamefont {Struzhkin}, \citenamefont {Moore}, \citenamefont {Levitas},
  \citenamefont {Jeanloz},\ and\ \citenamefont {Yao}}]{Hsieh2019}%
  \BibitemOpen
  \bibfield  {author} {\bibinfo {author} {\bibfnamefont {S.}~\bibnamefont
  {Hsieh}}, \bibinfo {author} {\bibfnamefont {P.}~\bibnamefont
  {Bhattacharyya}}, \bibinfo {author} {\bibfnamefont {C.}~\bibnamefont {Zu}},
  \bibinfo {author} {\bibfnamefont {T.}~\bibnamefont {Mittiga}}, \bibinfo
  {author} {\bibfnamefont {T.~J.}\ \bibnamefont {Smart}}, \bibinfo {author}
  {\bibfnamefont {F.}~\bibnamefont {Machado}}, \bibinfo {author} {\bibfnamefont
  {B.}~\bibnamefont {Kobrin}}, \bibinfo {author} {\bibfnamefont {T.~O.}\
  \bibnamefont {H{\"o}hn}}, \bibinfo {author} {\bibfnamefont {N.~Z.}\
  \bibnamefont {Rui}}, \bibinfo {author} {\bibfnamefont {M.}~\bibnamefont
  {Kamrani}}, \bibinfo {author} {\bibfnamefont {S.}~\bibnamefont {Chatterjee}},
  \bibinfo {author} {\bibfnamefont {S.}~\bibnamefont {Choi}}, \bibinfo {author}
  {\bibfnamefont {M.}~\bibnamefont {Zaletel}}, \bibinfo {author} {\bibfnamefont
  {V.~V.}\ \bibnamefont {Struzhkin}}, \bibinfo {author} {\bibfnamefont {J.~E.}\
  \bibnamefont {Moore}}, \bibinfo {author} {\bibfnamefont {V.~I.}\ \bibnamefont
  {Levitas}}, \bibinfo {author} {\bibfnamefont {R.}~\bibnamefont {Jeanloz}},\
  and\ \bibinfo {author} {\bibfnamefont {N.~Y.}\ \bibnamefont {Yao}},\
  }\bibfield  {title} {\bibinfo {title} {Imaging stress and magnetism at high
  pressures using a nanoscale quantum sensor},\ }\href
  {https://doi.org/10.1126/science.aaw4352} {\bibfield  {journal} {\bibinfo
  {journal} {Science}\ }\textbf {\bibinfo {volume} {366}},\ \bibinfo {pages}
  {1349} (\bibinfo {year} {2019})},\ \Eprint
  {https://arxiv.org/abs/https://science.sciencemag.org/content/366/6471/1349.full.pdf}
  {https://science.sciencemag.org/content/366/6471/1349.full.pdf} \BibitemShut
  {NoStop}%
\bibitem [{\citenamefont {Kucsko}\ \emph {et~al.}(2013)\citenamefont {Kucsko},
  \citenamefont {Maurer}, \citenamefont {Yao}, \citenamefont {Kubo},
  \citenamefont {Noh}, \citenamefont {Lo}, \citenamefont {Park},\ and\
  \citenamefont {Lukin}}]{Kucsko2013}%
  \BibitemOpen
  \bibfield  {author} {\bibinfo {author} {\bibfnamefont {G.}~\bibnamefont
  {Kucsko}}, \bibinfo {author} {\bibfnamefont {P.~C.}\ \bibnamefont {Maurer}},
  \bibinfo {author} {\bibfnamefont {N.~Y.}\ \bibnamefont {Yao}}, \bibinfo
  {author} {\bibfnamefont {M.}~\bibnamefont {Kubo}}, \bibinfo {author}
  {\bibfnamefont {H.~J.}\ \bibnamefont {Noh}}, \bibinfo {author} {\bibfnamefont
  {P.~K.}\ \bibnamefont {Lo}}, \bibinfo {author} {\bibfnamefont
  {H.}~\bibnamefont {Park}},\ and\ \bibinfo {author} {\bibfnamefont {M.~D.}\
  \bibnamefont {Lukin}},\ }\bibfield  {title} {\bibinfo {title}
  {Nanometre-scale thermometry in a living cell},\ }\href
  {https://doi.org/10.1038/nature12373} {\bibfield  {journal} {\bibinfo
  {journal} {Nature}\ }\textbf {\bibinfo {volume} {500}},\ \bibinfo {pages}
  {54} (\bibinfo {year} {2013})}\BibitemShut {NoStop}%
\bibitem [{\citenamefont {Dolde}\ \emph {et~al.}(2011)\citenamefont {Dolde},
  \citenamefont {Fedder}, \citenamefont {Doherty}, \citenamefont {N{\"o}bauer},
  \citenamefont {Rempp}, \citenamefont {Balasubramanian}, \citenamefont {Wolf},
  \citenamefont {Reinhard}, \citenamefont {Hollenberg}, \citenamefont
  {Jelezko},\ and\ \citenamefont {Wrachtrup}}]{Dolde2011}%
  \BibitemOpen
  \bibfield  {author} {\bibinfo {author} {\bibfnamefont {F.}~\bibnamefont
  {Dolde}}, \bibinfo {author} {\bibfnamefont {H.}~\bibnamefont {Fedder}},
  \bibinfo {author} {\bibfnamefont {M.~W.}\ \bibnamefont {Doherty}}, \bibinfo
  {author} {\bibfnamefont {T.}~\bibnamefont {N{\"o}bauer}}, \bibinfo {author}
  {\bibfnamefont {F.}~\bibnamefont {Rempp}}, \bibinfo {author} {\bibfnamefont
  {G.}~\bibnamefont {Balasubramanian}}, \bibinfo {author} {\bibfnamefont
  {T.}~\bibnamefont {Wolf}}, \bibinfo {author} {\bibfnamefont {F.}~\bibnamefont
  {Reinhard}}, \bibinfo {author} {\bibfnamefont {L.~C.~L.}\ \bibnamefont
  {Hollenberg}}, \bibinfo {author} {\bibfnamefont {F.}~\bibnamefont
  {Jelezko}},\ and\ \bibinfo {author} {\bibfnamefont {J.}~\bibnamefont
  {Wrachtrup}},\ }\bibfield  {title} {\bibinfo {title} {Electric-field sensing
  using single diamond spins},\ }\href {https://doi.org/10.1038/nphys1969}
  {\bibfield  {journal} {\bibinfo  {journal} {Nature Physics}\ }\textbf
  {\bibinfo {volume} {7}},\ \bibinfo {pages} {459} (\bibinfo {year}
  {2011})}\BibitemShut {NoStop}%
\bibitem [{\citenamefont {Nusran}\ \emph {et~al.}(2018)\citenamefont {Nusran},
  \citenamefont {Joshi}, \citenamefont {Cho}, \citenamefont {Tanatar},
  \citenamefont {Meier}, \citenamefont {Bud'ko}, \citenamefont {Canfield},
  \citenamefont {Liu}, \citenamefont {Lograsso},\ and\ \citenamefont
  {Prozorov}}]{Nusran2018}%
  \BibitemOpen
  \bibfield  {author} {\bibinfo {author} {\bibfnamefont {N.~M.}\ \bibnamefont
  {Nusran}}, \bibinfo {author} {\bibfnamefont {K.~R.}\ \bibnamefont {Joshi}},
  \bibinfo {author} {\bibfnamefont {K.}~\bibnamefont {Cho}}, \bibinfo {author}
  {\bibfnamefont {M.~A.}\ \bibnamefont {Tanatar}}, \bibinfo {author}
  {\bibfnamefont {W.~R.}\ \bibnamefont {Meier}}, \bibinfo {author}
  {\bibfnamefont {S.~L.}\ \bibnamefont {Bud'ko}}, \bibinfo {author}
  {\bibfnamefont {P.~C.}\ \bibnamefont {Canfield}}, \bibinfo {author}
  {\bibfnamefont {Y.}~\bibnamefont {Liu}}, \bibinfo {author} {\bibfnamefont
  {T.~A.}\ \bibnamefont {Lograsso}},\ and\ \bibinfo {author} {\bibfnamefont
  {R.}~\bibnamefont {Prozorov}},\ }\bibfield  {title} {\bibinfo {title}
  {Spatially-resolved study of the meissner effect in superconductors using
  {NV}-centers-in-diamond optical magnetometry},\ }\href
  {https://doi.org/10.1088/1367-2630/aab47c} {\bibfield  {journal} {\bibinfo
  {journal} {New Journal of Physics}\ }\textbf {\bibinfo {volume} {20}},\
  \bibinfo {pages} {043010} (\bibinfo {year} {2018})}\BibitemShut {NoStop}%
\bibitem [{\citenamefont {Broadway}\ \emph {et~al.}(2019)\citenamefont
  {Broadway}, \citenamefont {Johnson}, \citenamefont {Barson}, \citenamefont
  {Lillie}, \citenamefont {Dontschuk}, \citenamefont {McCloskey}, \citenamefont
  {Tsai}, \citenamefont {Teraji}, \citenamefont {Simpson}, \citenamefont
  {Stacey}, \citenamefont {McCallum}, \citenamefont {Bradby}, \citenamefont
  {Doherty}, \citenamefont {Hollenberg},\ and\ \citenamefont
  {Tetienne}}]{Broadway2019}%
  \BibitemOpen
  \bibfield  {author} {\bibinfo {author} {\bibfnamefont {D.~A.}\ \bibnamefont
  {Broadway}}, \bibinfo {author} {\bibfnamefont {B.~C.}\ \bibnamefont
  {Johnson}}, \bibinfo {author} {\bibfnamefont {M.~S.~J.}\ \bibnamefont
  {Barson}}, \bibinfo {author} {\bibfnamefont {S.~E.}\ \bibnamefont {Lillie}},
  \bibinfo {author} {\bibfnamefont {N.}~\bibnamefont {Dontschuk}}, \bibinfo
  {author} {\bibfnamefont {D.~J.}\ \bibnamefont {McCloskey}}, \bibinfo {author}
  {\bibfnamefont {A.}~\bibnamefont {Tsai}}, \bibinfo {author} {\bibfnamefont
  {T.}~\bibnamefont {Teraji}}, \bibinfo {author} {\bibfnamefont {D.~A.}\
  \bibnamefont {Simpson}}, \bibinfo {author} {\bibfnamefont {A.}~\bibnamefont
  {Stacey}}, \bibinfo {author} {\bibfnamefont {J.~C.}\ \bibnamefont
  {McCallum}}, \bibinfo {author} {\bibfnamefont {J.~E.}\ \bibnamefont
  {Bradby}}, \bibinfo {author} {\bibfnamefont {M.~W.}\ \bibnamefont {Doherty}},
  \bibinfo {author} {\bibfnamefont {L.~C.~L.}\ \bibnamefont {Hollenberg}},\
  and\ \bibinfo {author} {\bibfnamefont {J.-P.}\ \bibnamefont {Tetienne}},\
  }\bibfield  {title} {\bibinfo {title} {Microscopic imaging of the stress
  tensor in diamond using in situ quantum sensors},\ }\href
  {https://doi.org/10.1021/acs.nanolett.9b01402} {\bibfield  {journal}
  {\bibinfo  {journal} {Nano Letters}\ }\textbf {\bibinfo {volume} {19}},\
  \bibinfo {pages} {4543} (\bibinfo {year} {2019})},\ \bibinfo {note} {pMID:
  31150580},\ \Eprint
  {https://arxiv.org/abs/https://doi.org/10.1021/acs.nanolett.9b01402}
  {https://doi.org/10.1021/acs.nanolett.9b01402} \BibitemShut {NoStop}%
\bibitem [{\citenamefont {Joshi}\ \emph {et~al.}(2019)\citenamefont {Joshi},
  \citenamefont {Nusran}, \citenamefont {Tanatar}, \citenamefont {Cho},
  \citenamefont {Meier}, \citenamefont {Bud'ko}, \citenamefont {Canfield},\
  and\ \citenamefont {Prozorov}}]{Joshi2019}%
  \BibitemOpen
  \bibfield  {author} {\bibinfo {author} {\bibfnamefont {K.}~\bibnamefont
  {Joshi}}, \bibinfo {author} {\bibfnamefont {N.}~\bibnamefont {Nusran}},
  \bibinfo {author} {\bibfnamefont {M.}~\bibnamefont {Tanatar}}, \bibinfo
  {author} {\bibfnamefont {K.}~\bibnamefont {Cho}}, \bibinfo {author}
  {\bibfnamefont {W.}~\bibnamefont {Meier}}, \bibinfo {author} {\bibfnamefont
  {S.}~\bibnamefont {Bud'ko}}, \bibinfo {author} {\bibfnamefont
  {P.}~\bibnamefont {Canfield}},\ and\ \bibinfo {author} {\bibfnamefont
  {R.}~\bibnamefont {Prozorov}},\ }\bibfield  {title} {\bibinfo {title}
  {Measuring the lower critical field of superconductors using nitrogen-vacancy
  centers in diamond optical magnetometry},\ }\href
  {https://doi.org/10.1103/PhysRevApplied.11.014035} {\bibfield  {journal}
  {\bibinfo  {journal} {Phys. Rev. Applied}\ }\textbf {\bibinfo {volume}
  {11}},\ \bibinfo {pages} {014035} (\bibinfo {year} {2019})}\BibitemShut
  {NoStop}%
\bibitem [{\citenamefont {Schlussel}\ \emph {et~al.}(2018)\citenamefont
  {Schlussel}, \citenamefont {Lenz}, \citenamefont {Rohner}, \citenamefont
  {Bar-Haim}, \citenamefont {Bougas}, \citenamefont {Groswasser}, \citenamefont
  {Kieschnick}, \citenamefont {Rozenberg}, \citenamefont {Thiel}, \citenamefont
  {Waxman}, \citenamefont {Meijer}, \citenamefont {Maletinsky}, \citenamefont
  {Budker},\ and\ \citenamefont {Folman}}]{Schlussel2018}%
  \BibitemOpen
  \bibfield  {author} {\bibinfo {author} {\bibfnamefont {Y.}~\bibnamefont
  {Schlussel}}, \bibinfo {author} {\bibfnamefont {T.}~\bibnamefont {Lenz}},
  \bibinfo {author} {\bibfnamefont {D.}~\bibnamefont {Rohner}}, \bibinfo
  {author} {\bibfnamefont {Y.}~\bibnamefont {Bar-Haim}}, \bibinfo {author}
  {\bibfnamefont {L.}~\bibnamefont {Bougas}}, \bibinfo {author} {\bibfnamefont
  {D.}~\bibnamefont {Groswasser}}, \bibinfo {author} {\bibfnamefont
  {M.}~\bibnamefont {Kieschnick}}, \bibinfo {author} {\bibfnamefont
  {E.}~\bibnamefont {Rozenberg}}, \bibinfo {author} {\bibfnamefont
  {L.}~\bibnamefont {Thiel}}, \bibinfo {author} {\bibfnamefont
  {A.}~\bibnamefont {Waxman}}, \bibinfo {author} {\bibfnamefont
  {J.}~\bibnamefont {Meijer}}, \bibinfo {author} {\bibfnamefont
  {P.}~\bibnamefont {Maletinsky}}, \bibinfo {author} {\bibfnamefont
  {D.}~\bibnamefont {Budker}},\ and\ \bibinfo {author} {\bibfnamefont
  {R.}~\bibnamefont {Folman}},\ }\bibfield  {title} {\bibinfo {title}
  {Wide-field imaging of superconductor vortices with electron spins in
  diamond},\ }\href {https://doi.org/10.1103/PhysRevApplied.10.034032}
  {\bibfield  {journal} {\bibinfo  {journal} {Phys. Rev. Applied}\ }\textbf
  {\bibinfo {volume} {10}},\ \bibinfo {pages} {034032} (\bibinfo {year}
  {2018})}\BibitemShut {NoStop}%
\bibitem [{\citenamefont {Thiel}\ \emph {et~al.}(2016)\citenamefont {Thiel},
  \citenamefont {Rohner}, \citenamefont {Ganzhorn}, \citenamefont {Appel},
  \citenamefont {Neu}, \citenamefont {M{\"u}ller}, \citenamefont {Kleiner},
  \citenamefont {Koelle},\ and\ \citenamefont {Maletinsky}}]{Thiel2016}%
  \BibitemOpen
  \bibfield  {author} {\bibinfo {author} {\bibfnamefont {L.}~\bibnamefont
  {Thiel}}, \bibinfo {author} {\bibfnamefont {D.}~\bibnamefont {Rohner}},
  \bibinfo {author} {\bibfnamefont {M.}~\bibnamefont {Ganzhorn}}, \bibinfo
  {author} {\bibfnamefont {P.}~\bibnamefont {Appel}}, \bibinfo {author}
  {\bibfnamefont {E.}~\bibnamefont {Neu}}, \bibinfo {author} {\bibfnamefont
  {B.}~\bibnamefont {M{\"u}ller}}, \bibinfo {author} {\bibfnamefont
  {R.}~\bibnamefont {Kleiner}}, \bibinfo {author} {\bibfnamefont
  {D.}~\bibnamefont {Koelle}},\ and\ \bibinfo {author} {\bibfnamefont
  {P.}~\bibnamefont {Maletinsky}},\ }\bibfield  {title} {\bibinfo {title}
  {Quantitative nanoscale vortex imaging using a cryogenic quantum
  magnetometer},\ }\href {https://doi.org/10.1038/nnano.2016.63} {\bibfield
  {journal} {\bibinfo  {journal} {Nature Nanotechnology}\ }\textbf {\bibinfo
  {volume} {11}},\ \bibinfo {pages} {677} (\bibinfo {year} {2016})}\BibitemShut
  {NoStop}%
\bibitem [{\citenamefont {Steele}\ \emph {et~al.}(2017)\citenamefont {Steele},
  \citenamefont {Lawson}, \citenamefont {Onyszczak}, \citenamefont {Bush},
  \citenamefont {Mei}, \citenamefont {Dioguardi}, \citenamefont {King},
  \citenamefont {Parker}, \citenamefont {Pines}, \citenamefont {Weir},
  \citenamefont {Evans}, \citenamefont {Visbeck}, \citenamefont {Vohra},\ and\
  \citenamefont {Curro}}]{Steele2017}%
  \BibitemOpen
  \bibfield  {author} {\bibinfo {author} {\bibfnamefont {L.~G.}\ \bibnamefont
  {Steele}}, \bibinfo {author} {\bibfnamefont {M.}~\bibnamefont {Lawson}},
  \bibinfo {author} {\bibfnamefont {M.}~\bibnamefont {Onyszczak}}, \bibinfo
  {author} {\bibfnamefont {B.~T.}\ \bibnamefont {Bush}}, \bibinfo {author}
  {\bibfnamefont {Z.}~\bibnamefont {Mei}}, \bibinfo {author} {\bibfnamefont
  {A.~P.}\ \bibnamefont {Dioguardi}}, \bibinfo {author} {\bibfnamefont
  {J.}~\bibnamefont {King}}, \bibinfo {author} {\bibfnamefont {A.}~\bibnamefont
  {Parker}}, \bibinfo {author} {\bibfnamefont {A.}~\bibnamefont {Pines}},
  \bibinfo {author} {\bibfnamefont {S.~T.}\ \bibnamefont {Weir}}, \bibinfo
  {author} {\bibfnamefont {W.}~\bibnamefont {Evans}}, \bibinfo {author}
  {\bibfnamefont {K.}~\bibnamefont {Visbeck}}, \bibinfo {author} {\bibfnamefont
  {Y.~K.}\ \bibnamefont {Vohra}},\ and\ \bibinfo {author} {\bibfnamefont
  {N.~J.}\ \bibnamefont {Curro}},\ }\bibfield  {title} {\bibinfo {title}
  {Optically detected magnetic resonance of nitrogen vacancies in a diamond
  anvil cell using designer diamond anvils},\ }\href
  {https://doi.org/10.1063/1.5004153} {\bibfield  {journal} {\bibinfo
  {journal} {Applied Physics Letters}\ }\textbf {\bibinfo {volume} {111}},\
  \bibinfo {pages} {221903} (\bibinfo {year} {2017})},\ \Eprint
  {https://arxiv.org/abs/https://doi.org/10.1063/1.5004153}
  {https://doi.org/10.1063/1.5004153} \BibitemShut {NoStop}%
\bibitem [{\citenamefont {Neumann}\ \emph {et~al.}(2013)\citenamefont
  {Neumann}, \citenamefont {Jakobi}, \citenamefont {Dolde}, \citenamefont
  {Burk}, \citenamefont {Reuter}, \citenamefont {Waldherr}, \citenamefont
  {Honert}, \citenamefont {Wolf}, \citenamefont {Brunner}, \citenamefont
  {Shim}, \citenamefont {Suter}, \citenamefont {Sumiya}, \citenamefont
  {Isoya},\ and\ \citenamefont {Wrachtrup}}]{Neumann2013}%
  \BibitemOpen
  \bibfield  {author} {\bibinfo {author} {\bibfnamefont {P.}~\bibnamefont
  {Neumann}}, \bibinfo {author} {\bibfnamefont {I.}~\bibnamefont {Jakobi}},
  \bibinfo {author} {\bibfnamefont {F.}~\bibnamefont {Dolde}}, \bibinfo
  {author} {\bibfnamefont {C.}~\bibnamefont {Burk}}, \bibinfo {author}
  {\bibfnamefont {R.}~\bibnamefont {Reuter}}, \bibinfo {author} {\bibfnamefont
  {G.}~\bibnamefont {Waldherr}}, \bibinfo {author} {\bibfnamefont
  {J.}~\bibnamefont {Honert}}, \bibinfo {author} {\bibfnamefont
  {T.}~\bibnamefont {Wolf}}, \bibinfo {author} {\bibfnamefont {A.}~\bibnamefont
  {Brunner}}, \bibinfo {author} {\bibfnamefont {J.~H.}\ \bibnamefont {Shim}},
  \bibinfo {author} {\bibfnamefont {D.}~\bibnamefont {Suter}}, \bibinfo
  {author} {\bibfnamefont {H.}~\bibnamefont {Sumiya}}, \bibinfo {author}
  {\bibfnamefont {J.}~\bibnamefont {Isoya}},\ and\ \bibinfo {author}
  {\bibfnamefont {J.}~\bibnamefont {Wrachtrup}},\ }\bibfield  {title} {\bibinfo
  {title} {High-precision nanoscale temperature sensing using single defects in
  diamond},\ }\href {https://doi.org/10.1021/nl401216y} {\bibfield  {journal}
  {\bibinfo  {journal} {Nano Letters}\ }\textbf {\bibinfo {volume} {13}},\
  \bibinfo {pages} {2738} (\bibinfo {year} {2013})},\ \bibinfo {note} {pMID:
  23721106},\ \Eprint {https://arxiv.org/abs/https://doi.org/10.1021/nl401216y}
  {https://doi.org/10.1021/nl401216y} \BibitemShut {NoStop}%
\bibitem [{\citenamefont {Maurer}\ \emph {et~al.}(2012)\citenamefont {Maurer},
  \citenamefont {Kucsko}, \citenamefont {Latta}, \citenamefont {Jiang},
  \citenamefont {Yao}, \citenamefont {Bennett}, \citenamefont {Pastawski},
  \citenamefont {Hunger}, \citenamefont {Chisholm}, \citenamefont {Markham},
  \citenamefont {Twitchen}, \citenamefont {Cirac},\ and\ \citenamefont
  {Lukin}}]{Maurer2012}%
  \BibitemOpen
  \bibfield  {author} {\bibinfo {author} {\bibfnamefont {P.~C.}\ \bibnamefont
  {Maurer}}, \bibinfo {author} {\bibfnamefont {G.}~\bibnamefont {Kucsko}},
  \bibinfo {author} {\bibfnamefont {C.}~\bibnamefont {Latta}}, \bibinfo
  {author} {\bibfnamefont {L.}~\bibnamefont {Jiang}}, \bibinfo {author}
  {\bibfnamefont {N.~Y.}\ \bibnamefont {Yao}}, \bibinfo {author} {\bibfnamefont
  {S.~D.}\ \bibnamefont {Bennett}}, \bibinfo {author} {\bibfnamefont
  {F.}~\bibnamefont {Pastawski}}, \bibinfo {author} {\bibfnamefont
  {D.}~\bibnamefont {Hunger}}, \bibinfo {author} {\bibfnamefont
  {N.}~\bibnamefont {Chisholm}}, \bibinfo {author} {\bibfnamefont
  {M.}~\bibnamefont {Markham}}, \bibinfo {author} {\bibfnamefont {D.~J.}\
  \bibnamefont {Twitchen}}, \bibinfo {author} {\bibfnamefont {J.~I.}\
  \bibnamefont {Cirac}},\ and\ \bibinfo {author} {\bibfnamefont {M.~D.}\
  \bibnamefont {Lukin}},\ }\bibfield  {title} {\bibinfo {title}
  {Room-temperature quantum bit memory exceeding one second},\ }\href
  {https://doi.org/10.1126/science.1220513} {\bibfield  {journal} {\bibinfo
  {journal} {Science}\ }\textbf {\bibinfo {volume} {336}},\ \bibinfo {pages}
  {1283} (\bibinfo {year} {2012})},\ \Eprint
  {https://arxiv.org/abs/https://science.sciencemag.org/content/336/6086/1283.full.pdf}
  {https://science.sciencemag.org/content/336/6086/1283.full.pdf} \BibitemShut
  {NoStop}%
\bibitem [{\citenamefont {Bradley}\ \emph {et~al.}(2019)\citenamefont
  {Bradley}, \citenamefont {Randall}, \citenamefont {Abobeih}, \citenamefont
  {Berrevoets}, \citenamefont {Degen}, \citenamefont {Bakker}, \citenamefont
  {Markham}, \citenamefont {Twitchen},\ and\ \citenamefont
  {Taminiau}}]{Bradley2019}%
  \BibitemOpen
  \bibfield  {author} {\bibinfo {author} {\bibfnamefont {C.~E.}\ \bibnamefont
  {Bradley}}, \bibinfo {author} {\bibfnamefont {J.}~\bibnamefont {Randall}},
  \bibinfo {author} {\bibfnamefont {M.~H.}\ \bibnamefont {Abobeih}}, \bibinfo
  {author} {\bibfnamefont {R.~C.}\ \bibnamefont {Berrevoets}}, \bibinfo
  {author} {\bibfnamefont {M.~J.}\ \bibnamefont {Degen}}, \bibinfo {author}
  {\bibfnamefont {M.~A.}\ \bibnamefont {Bakker}}, \bibinfo {author}
  {\bibfnamefont {M.}~\bibnamefont {Markham}}, \bibinfo {author} {\bibfnamefont
  {D.~J.}\ \bibnamefont {Twitchen}},\ and\ \bibinfo {author} {\bibfnamefont
  {T.~H.}\ \bibnamefont {Taminiau}},\ }\bibfield  {title} {\bibinfo {title} {A
  ten-qubit solid-state spin register with quantum memory up to one minute},\
  }\href {https://doi.org/10.1103/PhysRevX.9.031045} {\bibfield  {journal}
  {\bibinfo  {journal} {Phys. Rev. X}\ }\textbf {\bibinfo {volume} {9}},\
  \bibinfo {pages} {031045} (\bibinfo {year} {2019})}\BibitemShut {NoStop}%
\bibitem [{\citenamefont {Cooper}\ \emph {et~al.}(2020)\citenamefont {Cooper},
  \citenamefont {Sun}, \citenamefont {Jaskula},\ and\ \citenamefont
  {Cappellaro}}]{Cooper2020}%
  \BibitemOpen
  \bibfield  {author} {\bibinfo {author} {\bibfnamefont {A.}~\bibnamefont
  {Cooper}}, \bibinfo {author} {\bibfnamefont {W.~K.~C.}\ \bibnamefont {Sun}},
  \bibinfo {author} {\bibfnamefont {J.-C.}\ \bibnamefont {Jaskula}},\ and\
  \bibinfo {author} {\bibfnamefont {P.}~\bibnamefont {Cappellaro}},\ }\bibfield
   {title} {\bibinfo {title} {Identification and control of electron-nuclear
  spin defects in diamond},\ }\href
  {https://doi.org/10.1103/PhysRevLett.124.083602} {\bibfield  {journal}
  {\bibinfo  {journal} {Phys. Rev. Lett.}\ }\textbf {\bibinfo {volume} {124}},\
  \bibinfo {pages} {083602} (\bibinfo {year} {2020})}\BibitemShut {NoStop}%
\bibitem [{\citenamefont {Yang}\ \emph {et~al.}(2016)\citenamefont {Yang},
  \citenamefont {Wang}, \citenamefont {Rao}, \citenamefont {Hien~Tran},
  \citenamefont {Momenzadeh}, \citenamefont {Markham}, \citenamefont
  {Twitchen}, \citenamefont {Wang}, \citenamefont {Yang}, \citenamefont
  {St{\"o}hr}, \citenamefont {Neumann}, \citenamefont {Kosaka},\ and\
  \citenamefont {Wrachtrup}}]{Yang2016}%
  \BibitemOpen
  \bibfield  {author} {\bibinfo {author} {\bibfnamefont {S.}~\bibnamefont
  {Yang}}, \bibinfo {author} {\bibfnamefont {Y.}~\bibnamefont {Wang}}, \bibinfo
  {author} {\bibfnamefont {D.~D.~B.}\ \bibnamefont {Rao}}, \bibinfo {author}
  {\bibfnamefont {T.}~\bibnamefont {Hien~Tran}}, \bibinfo {author}
  {\bibfnamefont {A.~S.}\ \bibnamefont {Momenzadeh}}, \bibinfo {author}
  {\bibfnamefont {M.}~\bibnamefont {Markham}}, \bibinfo {author} {\bibfnamefont
  {D.~J.}\ \bibnamefont {Twitchen}}, \bibinfo {author} {\bibfnamefont
  {P.}~\bibnamefont {Wang}}, \bibinfo {author} {\bibfnamefont {W.}~\bibnamefont
  {Yang}}, \bibinfo {author} {\bibfnamefont {R.}~\bibnamefont {St{\"o}hr}},
  \bibinfo {author} {\bibfnamefont {P.}~\bibnamefont {Neumann}}, \bibinfo
  {author} {\bibfnamefont {H.}~\bibnamefont {Kosaka}},\ and\ \bibinfo {author}
  {\bibfnamefont {J.}~\bibnamefont {Wrachtrup}},\ }\bibfield  {title} {\bibinfo
  {title} {High-fidelity transfer and storage of photon states in a single
  nuclear spin},\ }\href {https://doi.org/10.1038/nphoton.2016.103} {\bibfield
  {journal} {\bibinfo  {journal} {Nature Photonics}\ }\textbf {\bibinfo
  {volume} {10}},\ \bibinfo {pages} {507} (\bibinfo {year} {2016})}\BibitemShut
  {NoStop}%
\bibitem [{\citenamefont {Hensen}\ \emph {et~al.}(2015)\citenamefont {Hensen},
  \citenamefont {Bernien}, \citenamefont {Dr{\'e}au}, \citenamefont {Reiserer},
  \citenamefont {Kalb}, \citenamefont {Blok}, \citenamefont {Ruitenberg},
  \citenamefont {Vermeulen}, \citenamefont {Schouten}, \citenamefont
  {Abell{\'a}n}, \citenamefont {Amaya}, \citenamefont {Pruneri}, \citenamefont
  {Mitchell}, \citenamefont {Markham}, \citenamefont {Twitchen}, \citenamefont
  {Elkouss}, \citenamefont {Wehner}, \citenamefont {Taminiau},\ and\
  \citenamefont {Hanson}}]{Hensen2015}%
  \BibitemOpen
  \bibfield  {author} {\bibinfo {author} {\bibfnamefont {B.}~\bibnamefont
  {Hensen}}, \bibinfo {author} {\bibfnamefont {H.}~\bibnamefont {Bernien}},
  \bibinfo {author} {\bibfnamefont {A.~E.}\ \bibnamefont {Dr{\'e}au}}, \bibinfo
  {author} {\bibfnamefont {A.}~\bibnamefont {Reiserer}}, \bibinfo {author}
  {\bibfnamefont {N.}~\bibnamefont {Kalb}}, \bibinfo {author} {\bibfnamefont
  {M.~S.}\ \bibnamefont {Blok}}, \bibinfo {author} {\bibfnamefont
  {J.}~\bibnamefont {Ruitenberg}}, \bibinfo {author} {\bibfnamefont {R.~F.~L.}\
  \bibnamefont {Vermeulen}}, \bibinfo {author} {\bibfnamefont {R.~N.}\
  \bibnamefont {Schouten}}, \bibinfo {author} {\bibfnamefont {C.}~\bibnamefont
  {Abell{\'a}n}}, \bibinfo {author} {\bibfnamefont {W.}~\bibnamefont {Amaya}},
  \bibinfo {author} {\bibfnamefont {V.}~\bibnamefont {Pruneri}}, \bibinfo
  {author} {\bibfnamefont {M.~W.}\ \bibnamefont {Mitchell}}, \bibinfo {author}
  {\bibfnamefont {M.}~\bibnamefont {Markham}}, \bibinfo {author} {\bibfnamefont
  {D.~J.}\ \bibnamefont {Twitchen}}, \bibinfo {author} {\bibfnamefont
  {D.}~\bibnamefont {Elkouss}}, \bibinfo {author} {\bibfnamefont
  {S.}~\bibnamefont {Wehner}}, \bibinfo {author} {\bibfnamefont {T.~H.}\
  \bibnamefont {Taminiau}},\ and\ \bibinfo {author} {\bibfnamefont
  {R.}~\bibnamefont {Hanson}},\ }\bibfield  {title} {\bibinfo {title}
  {Loophole-free bell inequality violation using electron spins separated by
  1.3 kilometres},\ }\href {https://doi.org/10.1038/nature15759} {\bibfield
  {journal} {\bibinfo  {journal} {Nature}\ }\textbf {\bibinfo {volume} {526}},\
  \bibinfo {pages} {682} (\bibinfo {year} {2015})}\BibitemShut {NoStop}%
\bibitem [{\citenamefont {Neumann}\ \emph {et~al.}(2008)\citenamefont
  {Neumann}, \citenamefont {Mizuochi}, \citenamefont {Rempp}, \citenamefont
  {Hemmer}, \citenamefont {Watanabe}, \citenamefont {Yamasaki}, \citenamefont
  {Jacques}, \citenamefont {Gaebel}, \citenamefont {Jelezko},\ and\
  \citenamefont {Wrachtrup}}]{Neumann2008}%
  \BibitemOpen
  \bibfield  {author} {\bibinfo {author} {\bibfnamefont {P.}~\bibnamefont
  {Neumann}}, \bibinfo {author} {\bibfnamefont {N.}~\bibnamefont {Mizuochi}},
  \bibinfo {author} {\bibfnamefont {F.}~\bibnamefont {Rempp}}, \bibinfo
  {author} {\bibfnamefont {P.}~\bibnamefont {Hemmer}}, \bibinfo {author}
  {\bibfnamefont {H.}~\bibnamefont {Watanabe}}, \bibinfo {author}
  {\bibfnamefont {S.}~\bibnamefont {Yamasaki}}, \bibinfo {author}
  {\bibfnamefont {V.}~\bibnamefont {Jacques}}, \bibinfo {author} {\bibfnamefont
  {T.}~\bibnamefont {Gaebel}}, \bibinfo {author} {\bibfnamefont
  {F.}~\bibnamefont {Jelezko}},\ and\ \bibinfo {author} {\bibfnamefont
  {J.}~\bibnamefont {Wrachtrup}},\ }\bibfield  {title} {\bibinfo {title}
  {Multipartite entanglement among single spins in diamond},\ }\href
  {https://doi.org/10.1126/science.1157233} {\bibfield  {journal} {\bibinfo
  {journal} {Science}\ }\textbf {\bibinfo {volume} {320}},\ \bibinfo {pages}
  {1326} (\bibinfo {year} {2008})},\ \Eprint
  {https://arxiv.org/abs/https://science.sciencemag.org/content/320/5881/1326.full.pdf}
  {https://science.sciencemag.org/content/320/5881/1326.full.pdf} \BibitemShut
  {NoStop}%
\bibitem [{\citenamefont {Doherty}\ \emph {et~al.}(2013)\citenamefont
  {Doherty}, \citenamefont {Manson}, \citenamefont {Delaney}, \citenamefont
  {Jelezko}, \citenamefont {Wrachtrup},\ and\ \citenamefont
  {Hollenberg}}]{Doherty2013}%
  \BibitemOpen
  \bibfield  {author} {\bibinfo {author} {\bibfnamefont {M.~W.}\ \bibnamefont
  {Doherty}}, \bibinfo {author} {\bibfnamefont {N.~B.}\ \bibnamefont {Manson}},
  \bibinfo {author} {\bibfnamefont {P.}~\bibnamefont {Delaney}}, \bibinfo
  {author} {\bibfnamefont {F.}~\bibnamefont {Jelezko}}, \bibinfo {author}
  {\bibfnamefont {J.}~\bibnamefont {Wrachtrup}},\ and\ \bibinfo {author}
  {\bibfnamefont {L.~C.}\ \bibnamefont {Hollenberg}},\ }\bibfield  {title}
  {\bibinfo {title} {The nitrogen-vacancy colour centre in diamond},\ }\href
  {https://doi.org/https://doi.org/10.1016/j.physrep.2013.02.001} {\bibfield
  {journal} {\bibinfo  {journal} {Physics Reports}\ }\textbf {\bibinfo {volume}
  {528}},\ \bibinfo {pages} {1} (\bibinfo {year} {2013})},\ \bibinfo {note}
  {the nitrogen-vacancy colour centre in diamond}\BibitemShut {NoStop}%
\bibitem [{\citenamefont {Ádám Gali}(2019)}]{Gali2019Ab}%
  \BibitemOpen
  \bibfield  {author} {\bibinfo {author} {\bibnamefont {Ádám Gali}},\
  }\bibfield  {title} {\bibinfo {title} {Ab initio theory of the
  nitrogen-vacancy center in diamond},\ }\href
  {https://doi.org/doi:10.1515/nanoph-2019-0154} {\bibfield  {journal}
  {\bibinfo  {journal} {Nanophotonics}\ }\textbf {\bibinfo {volume} {8}},\
  \bibinfo {pages} {1907} (\bibinfo {year} {2019})}\BibitemShut {NoStop}%
\bibitem [{\citenamefont {Doherty}\ \emph {et~al.}(2012)\citenamefont
  {Doherty}, \citenamefont {Dolde}, \citenamefont {Fedder}, \citenamefont
  {Jelezko}, \citenamefont {Wrachtrup}, \citenamefont {Manson},\ and\
  \citenamefont {Hollenberg}}]{Doherty2012Theory}%
  \BibitemOpen
  \bibfield  {author} {\bibinfo {author} {\bibfnamefont {M.~W.}\ \bibnamefont
  {Doherty}}, \bibinfo {author} {\bibfnamefont {F.}~\bibnamefont {Dolde}},
  \bibinfo {author} {\bibfnamefont {H.}~\bibnamefont {Fedder}}, \bibinfo
  {author} {\bibfnamefont {F.}~\bibnamefont {Jelezko}}, \bibinfo {author}
  {\bibfnamefont {J.}~\bibnamefont {Wrachtrup}}, \bibinfo {author}
  {\bibfnamefont {N.~B.}\ \bibnamefont {Manson}},\ and\ \bibinfo {author}
  {\bibfnamefont {L.~C.~L.}\ \bibnamefont {Hollenberg}},\ }\bibfield  {title}
  {\bibinfo {title} {Theory of the ground-state spin of the
  nv${}^{\ensuremath{-}}$ center in diamond},\ }\href
  {https://doi.org/10.1103/PhysRevB.85.205203} {\bibfield  {journal} {\bibinfo
  {journal} {Phys. Rev. B}\ }\textbf {\bibinfo {volume} {85}},\ \bibinfo
  {pages} {205203} (\bibinfo {year} {2012})}\BibitemShut {NoStop}%
\bibitem [{\citenamefont {Falk}\ \emph {et~al.}(2014)\citenamefont {Falk},
  \citenamefont {Klimov}, \citenamefont {Buckley}, \citenamefont {Iv\'ady},
  \citenamefont {Abrikosov}, \citenamefont {Calusine}, \citenamefont {Koehl},
  \citenamefont {Gali},\ and\ \citenamefont
  {Awschalom}}]{Falk2014Electrically}%
  \BibitemOpen
  \bibfield  {author} {\bibinfo {author} {\bibfnamefont {A.~L.}\ \bibnamefont
  {Falk}}, \bibinfo {author} {\bibfnamefont {P.~V.}\ \bibnamefont {Klimov}},
  \bibinfo {author} {\bibfnamefont {B.~B.}\ \bibnamefont {Buckley}}, \bibinfo
  {author} {\bibfnamefont {V.}~\bibnamefont {Iv\'ady}}, \bibinfo {author}
  {\bibfnamefont {I.~A.}\ \bibnamefont {Abrikosov}}, \bibinfo {author}
  {\bibfnamefont {G.}~\bibnamefont {Calusine}}, \bibinfo {author}
  {\bibfnamefont {W.~F.}\ \bibnamefont {Koehl}}, \bibinfo {author}
  {\bibfnamefont {A.}~\bibnamefont {Gali}},\ and\ \bibinfo {author}
  {\bibfnamefont {D.~D.}\ \bibnamefont {Awschalom}},\ }\bibfield  {title}
  {\bibinfo {title} {Electrically and mechanically tunable electron spins in
  silicon carbide color centers},\ }\href
  {https://doi.org/10.1103/PhysRevLett.112.187601} {\bibfield  {journal}
  {\bibinfo  {journal} {Phys. Rev. Lett.}\ }\textbf {\bibinfo {volume} {112}},\
  \bibinfo {pages} {187601} (\bibinfo {year} {2014})}\BibitemShut {NoStop}%
\bibitem [{\citenamefont {Udvarhelyi}\ \emph {et~al.}(2018)\citenamefont
  {Udvarhelyi}, \citenamefont {Shkolnikov}, \citenamefont {Gali}, \citenamefont
  {Burkard},\ and\ \citenamefont {P\'alyi}}]{Udvarhelyi2018Spin}%
  \BibitemOpen
  \bibfield  {author} {\bibinfo {author} {\bibfnamefont {P.}~\bibnamefont
  {Udvarhelyi}}, \bibinfo {author} {\bibfnamefont {V.~O.}\ \bibnamefont
  {Shkolnikov}}, \bibinfo {author} {\bibfnamefont {A.}~\bibnamefont {Gali}},
  \bibinfo {author} {\bibfnamefont {G.}~\bibnamefont {Burkard}},\ and\ \bibinfo
  {author} {\bibfnamefont {A.}~\bibnamefont {P\'alyi}},\ }\bibfield  {title}
  {\bibinfo {title} {Spin-strain interaction in nitrogen-vacancy centers in
  diamond},\ }\href {https://doi.org/10.1103/PhysRevB.98.075201} {\bibfield
  {journal} {\bibinfo  {journal} {Phys. Rev. B}\ }\textbf {\bibinfo {volume}
  {98}},\ \bibinfo {pages} {075201} (\bibinfo {year} {2018})}\BibitemShut
  {NoStop}%
\bibitem [{\citenamefont {Maze}\ \emph {et~al.}(2011)\citenamefont {Maze},
  \citenamefont {Gali}, \citenamefont {Togan}, \citenamefont {Chu},
  \citenamefont {Trifonov}, \citenamefont {Kaxiras},\ and\ \citenamefont
  {Lukin}}]{Maze2011Properties}%
  \BibitemOpen
  \bibfield  {author} {\bibinfo {author} {\bibfnamefont {J.~R.}\ \bibnamefont
  {Maze}}, \bibinfo {author} {\bibfnamefont {A.}~\bibnamefont {Gali}}, \bibinfo
  {author} {\bibfnamefont {E.}~\bibnamefont {Togan}}, \bibinfo {author}
  {\bibfnamefont {Y.}~\bibnamefont {Chu}}, \bibinfo {author} {\bibfnamefont
  {A.}~\bibnamefont {Trifonov}}, \bibinfo {author} {\bibfnamefont
  {E.}~\bibnamefont {Kaxiras}},\ and\ \bibinfo {author} {\bibfnamefont {M.~D.}\
  \bibnamefont {Lukin}},\ }\bibfield  {title} {\bibinfo {title} {Properties of
  nitrogen-vacancy centers in diamond: the group theoretic approach},\ }\href
  {https://doi.org/10.1088/1367-2630/13/2/025025} {\bibfield  {journal}
  {\bibinfo  {journal} {New Journal of Physics}\ }\textbf {\bibinfo {volume}
  {13}},\ \bibinfo {pages} {025025} (\bibinfo {year} {2011})}\BibitemShut
  {NoStop}%
\bibitem [{\citenamefont {Doherty}\ \emph {et~al.}(2011)\citenamefont
  {Doherty}, \citenamefont {Manson}, \citenamefont {Delaney},\ and\
  \citenamefont {Hollenberg}}]{Doherty2011The}%
  \BibitemOpen
  \bibfield  {author} {\bibinfo {author} {\bibfnamefont {M.~W.}\ \bibnamefont
  {Doherty}}, \bibinfo {author} {\bibfnamefont {N.~B.}\ \bibnamefont {Manson}},
  \bibinfo {author} {\bibfnamefont {P.}~\bibnamefont {Delaney}},\ and\ \bibinfo
  {author} {\bibfnamefont {L.~C.~L.}\ \bibnamefont {Hollenberg}},\ }\bibfield
  {title} {\bibinfo {title} {The negatively charged nitrogen-vacancy centre in
  diamond: the electronic solution},\ }\href
  {https://doi.org/10.1088/1367-2630/13/2/025019} {\bibfield  {journal}
  {\bibinfo  {journal} {New Journal of Physics}\ }\textbf {\bibinfo {volume}
  {13}},\ \bibinfo {pages} {025019} (\bibinfo {year} {2011})}\BibitemShut
  {NoStop}%
\bibitem [{\citenamefont {Doherty}\ \emph
  {et~al.}(2014{\natexlab{b}})\citenamefont {Doherty}, \citenamefont {Acosta},
  \citenamefont {Jarmola}, \citenamefont {Barson}, \citenamefont {Manson},
  \citenamefont {Budker},\ and\ \citenamefont
  {Hollenberg}}]{Doherty2014Temperature}%
  \BibitemOpen
  \bibfield  {author} {\bibinfo {author} {\bibfnamefont {M.~W.}\ \bibnamefont
  {Doherty}}, \bibinfo {author} {\bibfnamefont {V.~M.}\ \bibnamefont {Acosta}},
  \bibinfo {author} {\bibfnamefont {A.}~\bibnamefont {Jarmola}}, \bibinfo
  {author} {\bibfnamefont {M.~S.~J.}\ \bibnamefont {Barson}}, \bibinfo {author}
  {\bibfnamefont {N.~B.}\ \bibnamefont {Manson}}, \bibinfo {author}
  {\bibfnamefont {D.}~\bibnamefont {Budker}},\ and\ \bibinfo {author}
  {\bibfnamefont {L.~C.~L.}\ \bibnamefont {Hollenberg}},\ }\bibfield  {title}
  {\bibinfo {title} {Temperature shifts of the resonances of the
  ${\mathrm{nv}}^{\ensuremath{-}}$ center in diamond},\ }\href
  {https://doi.org/10.1103/PhysRevB.90.041201} {\bibfield  {journal} {\bibinfo
  {journal} {Phys. Rev. B}\ }\textbf {\bibinfo {volume} {90}},\ \bibinfo
  {pages} {041201} (\bibinfo {year} {2014}{\natexlab{b}})}\BibitemShut
  {NoStop}%
\bibitem [{\citenamefont {Iv\'ady}\ \emph {et~al.}(2014)\citenamefont
  {Iv\'ady}, \citenamefont {Simon}, \citenamefont {Maze}, \citenamefont
  {Abrikosov},\ and\ \citenamefont {Gali}}]{Ivady2014Pressure}%
  \BibitemOpen
  \bibfield  {author} {\bibinfo {author} {\bibfnamefont {V.}~\bibnamefont
  {Iv\'ady}}, \bibinfo {author} {\bibfnamefont {T.}~\bibnamefont {Simon}},
  \bibinfo {author} {\bibfnamefont {J.~R.}\ \bibnamefont {Maze}}, \bibinfo
  {author} {\bibfnamefont {I.~A.}\ \bibnamefont {Abrikosov}},\ and\ \bibinfo
  {author} {\bibfnamefont {A.}~\bibnamefont {Gali}},\ }\bibfield  {title}
  {\bibinfo {title} {Pressure and temperature dependence of the zero-field
  splitting in the ground state of nv centers in diamond: A first-principles
  study},\ }\href {https://doi.org/10.1103/PhysRevB.90.235205} {\bibfield
  {journal} {\bibinfo  {journal} {Phys. Rev. B}\ }\textbf {\bibinfo {volume}
  {90}},\ \bibinfo {pages} {235205} (\bibinfo {year} {2014})}\BibitemShut
  {NoStop}%
\bibitem [{\citenamefont {Kobayashi}\ and\ \citenamefont
  {Nisida}(1993)}]{Kobayashi1993High}%
  \BibitemOpen
  \bibfield  {author} {\bibinfo {author} {\bibfnamefont {M.}~\bibnamefont
  {Kobayashi}}\ and\ \bibinfo {author} {\bibfnamefont {Y.}~\bibnamefont
  {Nisida}},\ }\bibfield  {title} {\bibinfo {title} {High pressure effects on
  photoluminescence spectra of color centers in diamond},\ }\href
  {https://doi.org/10.7567/jjaps.32s1.279} {\bibfield  {journal} {\bibinfo
  {journal} {Japanese Journal of Applied Physics}\ }\textbf {\bibinfo {volume}
  {32}},\ \bibinfo {pages} {279} (\bibinfo {year} {1993})}\BibitemShut
  {NoStop}%
\bibitem [{\citenamefont {Deng}\ \emph {et~al.}(2014)\citenamefont {Deng},
  \citenamefont {Zhang},\ and\ \citenamefont {Shi}}]{Deng2014New}%
  \BibitemOpen
  \bibfield  {author} {\bibinfo {author} {\bibfnamefont {B.}~\bibnamefont
  {Deng}}, \bibinfo {author} {\bibfnamefont {R.~Q.}\ \bibnamefont {Zhang}},\
  and\ \bibinfo {author} {\bibfnamefont {X.~Q.}\ \bibnamefont {Shi}},\
  }\bibfield  {title} {\bibinfo {title} {New insight into the spin-conserving
  excitation of the negatively charged nitrogen-vacancy center in diamond},\
  }\href {https://doi.org/10.1038/srep05144} {\bibfield  {journal} {\bibinfo
  {journal} {Scientific Reports}\ }\textbf {\bibinfo {volume} {4}},\ \bibinfo
  {pages} {5144} (\bibinfo {year} {2014})}\BibitemShut {NoStop}%
\bibitem [{\citenamefont {Udvarhelyi}\ and\ \citenamefont
  {Gali}(2018)}]{Udvarhelyi2018Ab}%
  \BibitemOpen
  \bibfield  {author} {\bibinfo {author} {\bibfnamefont {P.}~\bibnamefont
  {Udvarhelyi}}\ and\ \bibinfo {author} {\bibfnamefont {A.}~\bibnamefont
  {Gali}},\ }\bibfield  {title} {\bibinfo {title} {Ab initio spin-strain
  coupling parameters of divacancy qubits in silicon carbide},\ }\href
  {https://doi.org/10.1103/PhysRevApplied.10.054010} {\bibfield  {journal}
  {\bibinfo  {journal} {Phys. Rev. Applied}\ }\textbf {\bibinfo {volume}
  {10}},\ \bibinfo {pages} {054010} (\bibinfo {year} {2018})}\BibitemShut
  {NoStop}%
\bibitem [{\citenamefont {Loubser}\ and\ \citenamefont {van
  Wyk}(1978)}]{Loubser1978}%
  \BibitemOpen
  \bibfield  {author} {\bibinfo {author} {\bibfnamefont {J.~H.~N.}\
  \bibnamefont {Loubser}}\ and\ \bibinfo {author} {\bibfnamefont {J.~A.}\
  \bibnamefont {van Wyk}},\ }\bibfield  {title} {\bibinfo {title} {Electron
  spin resonance in the study of diamond},\ }\href
  {https://doi.org/10.1088/0034-4885/41/8/002} {\bibfield  {journal} {\bibinfo
  {journal} {Reports on Progress in Physics}\ }\textbf {\bibinfo {volume}
  {41}},\ \bibinfo {pages} {1201} (\bibinfo {year} {1978})}\BibitemShut
  {NoStop}%
\bibitem [{\citenamefont {Simanovskaia}\ \emph {et~al.}(2013)\citenamefont
  {Simanovskaia}, \citenamefont {Jensen}, \citenamefont {Jarmola},
  \citenamefont {Aulenbacher}, \citenamefont {Manson},\ and\ \citenamefont
  {Budker}}]{Simanovskaia2013}%
  \BibitemOpen
  \bibfield  {author} {\bibinfo {author} {\bibfnamefont {M.}~\bibnamefont
  {Simanovskaia}}, \bibinfo {author} {\bibfnamefont {K.}~\bibnamefont
  {Jensen}}, \bibinfo {author} {\bibfnamefont {A.}~\bibnamefont {Jarmola}},
  \bibinfo {author} {\bibfnamefont {K.}~\bibnamefont {Aulenbacher}}, \bibinfo
  {author} {\bibfnamefont {N.}~\bibnamefont {Manson}},\ and\ \bibinfo {author}
  {\bibfnamefont {D.}~\bibnamefont {Budker}},\ }\bibfield  {title} {\bibinfo
  {title} {Sidebands in optically detected magnetic resonance signals of
  nitrogen vacancy centers in diamond},\ }\href
  {https://doi.org/10.1103/PhysRevB.87.224106} {\bibfield  {journal} {\bibinfo
  {journal} {Phys. Rev. B}\ }\textbf {\bibinfo {volume} {87}},\ \bibinfo
  {pages} {224106} (\bibinfo {year} {2013})}\BibitemShut {NoStop}%
\bibitem [{\citenamefont {Felton}\ \emph {et~al.}(2009)\citenamefont {Felton},
  \citenamefont {Edmonds}, \citenamefont {Newton}, \citenamefont {Martineau},
  \citenamefont {Fisher}, \citenamefont {Twitchen},\ and\ \citenamefont
  {Baker}}]{Felton2009}%
  \BibitemOpen
  \bibfield  {author} {\bibinfo {author} {\bibfnamefont {S.}~\bibnamefont
  {Felton}}, \bibinfo {author} {\bibfnamefont {A.~M.}\ \bibnamefont {Edmonds}},
  \bibinfo {author} {\bibfnamefont {M.~E.}\ \bibnamefont {Newton}}, \bibinfo
  {author} {\bibfnamefont {P.~M.}\ \bibnamefont {Martineau}}, \bibinfo {author}
  {\bibfnamefont {D.}~\bibnamefont {Fisher}}, \bibinfo {author} {\bibfnamefont
  {D.~J.}\ \bibnamefont {Twitchen}},\ and\ \bibinfo {author} {\bibfnamefont
  {J.~M.}\ \bibnamefont {Baker}},\ }\bibfield  {title} {\bibinfo {title}
  {Hyperfine interaction in the ground state of the negatively charged nitrogen
  vacancy center in diamond},\ }\href
  {https://doi.org/10.1103/PhysRevB.79.075203} {\bibfield  {journal} {\bibinfo
  {journal} {Phys. Rev. B}\ }\textbf {\bibinfo {volume} {79}},\ \bibinfo
  {pages} {075203} (\bibinfo {year} {2009})}\BibitemShut {NoStop}%
\bibitem [{\citenamefont {Nizovtsev}\ \emph
  {et~al.}(2010{\natexlab{a}})\citenamefont {Nizovtsev}, \citenamefont {Kilin},
  \citenamefont {Neumann}, \citenamefont {Jelezko},\ and\ \citenamefont
  {Wrachtrup}}]{Nizovtsev2010}%
  \BibitemOpen
  \bibfield  {author} {\bibinfo {author} {\bibfnamefont {A.~P.}\ \bibnamefont
  {Nizovtsev}}, \bibinfo {author} {\bibfnamefont {S.~Y.}\ \bibnamefont
  {Kilin}}, \bibinfo {author} {\bibfnamefont {P.}~\bibnamefont {Neumann}},
  \bibinfo {author} {\bibfnamefont {F.}~\bibnamefont {Jelezko}},\ and\ \bibinfo
  {author} {\bibfnamefont {J.}~\bibnamefont {Wrachtrup}},\ }\bibfield  {title}
  {\bibinfo {title} {Quantum registers based on single nv + n13c centers in
  diamond: Ii. spin characteristics of registers and spectra of optically
  detected magnetic resonance},\ }\href
  {https://doi.org/10.1134/S0030400X1002013X} {\bibfield  {journal} {\bibinfo
  {journal} {Optics and Spectroscopy}\ }\textbf {\bibinfo {volume} {108}},\
  \bibinfo {pages} {239} (\bibinfo {year} {2010}{\natexlab{a}})}\BibitemShut
  {NoStop}%
\bibitem [{\citenamefont {Mizuochi}\ \emph {et~al.}(2009)\citenamefont
  {Mizuochi}, \citenamefont {Neumann}, \citenamefont {Rempp}, \citenamefont
  {Beck}, \citenamefont {Jacques}, \citenamefont {Siyushev}, \citenamefont
  {Nakamura}, \citenamefont {Twitchen}, \citenamefont {Watanabe}, \citenamefont
  {Yamasaki}, \citenamefont {Jelezko},\ and\ \citenamefont
  {Wrachtrup}}]{Mizuochi2009}%
  \BibitemOpen
  \bibfield  {author} {\bibinfo {author} {\bibfnamefont {N.}~\bibnamefont
  {Mizuochi}}, \bibinfo {author} {\bibfnamefont {P.}~\bibnamefont {Neumann}},
  \bibinfo {author} {\bibfnamefont {F.}~\bibnamefont {Rempp}}, \bibinfo
  {author} {\bibfnamefont {J.}~\bibnamefont {Beck}}, \bibinfo {author}
  {\bibfnamefont {V.}~\bibnamefont {Jacques}}, \bibinfo {author} {\bibfnamefont
  {P.}~\bibnamefont {Siyushev}}, \bibinfo {author} {\bibfnamefont
  {K.}~\bibnamefont {Nakamura}}, \bibinfo {author} {\bibfnamefont {D.~J.}\
  \bibnamefont {Twitchen}}, \bibinfo {author} {\bibfnamefont {H.}~\bibnamefont
  {Watanabe}}, \bibinfo {author} {\bibfnamefont {S.}~\bibnamefont {Yamasaki}},
  \bibinfo {author} {\bibfnamefont {F.}~\bibnamefont {Jelezko}},\ and\ \bibinfo
  {author} {\bibfnamefont {J.}~\bibnamefont {Wrachtrup}},\ }\bibfield  {title}
  {\bibinfo {title} {Coherence of single spins coupled to a nuclear spin bath
  of varying density},\ }\href {https://doi.org/10.1103/PhysRevB.80.041201}
  {\bibfield  {journal} {\bibinfo  {journal} {Phys. Rev. B}\ }\textbf {\bibinfo
  {volume} {80}},\ \bibinfo {pages} {041201} (\bibinfo {year}
  {2009})}\BibitemShut {NoStop}%
\bibitem [{\citenamefont {Baranov}\ \emph {et~al.}(2011)\citenamefont
  {Baranov}, \citenamefont {Soltamova}, \citenamefont {Tolmachev},
  \citenamefont {Romanov}, \citenamefont {Babunts}, \citenamefont {Shakhov},
  \citenamefont {Kidalov}, \citenamefont {Vul’}, \citenamefont {Mamin},
  \citenamefont {Orlinskii},\ and\ \citenamefont {Silkin}}]{Baranov2011}%
  \BibitemOpen
  \bibfield  {author} {\bibinfo {author} {\bibfnamefont {P.~G.}\ \bibnamefont
  {Baranov}}, \bibinfo {author} {\bibfnamefont {A.~A.}\ \bibnamefont
  {Soltamova}}, \bibinfo {author} {\bibfnamefont {D.~O.}\ \bibnamefont
  {Tolmachev}}, \bibinfo {author} {\bibfnamefont {N.~G.}\ \bibnamefont
  {Romanov}}, \bibinfo {author} {\bibfnamefont {R.~A.}\ \bibnamefont
  {Babunts}}, \bibinfo {author} {\bibfnamefont {F.~M.}\ \bibnamefont
  {Shakhov}}, \bibinfo {author} {\bibfnamefont {S.~V.}\ \bibnamefont
  {Kidalov}}, \bibinfo {author} {\bibfnamefont {A.~Y.}\ \bibnamefont {Vul’}},
  \bibinfo {author} {\bibfnamefont {G.~V.}\ \bibnamefont {Mamin}}, \bibinfo
  {author} {\bibfnamefont {S.~B.}\ \bibnamefont {Orlinskii}},\ and\ \bibinfo
  {author} {\bibfnamefont {N.~I.}\ \bibnamefont {Silkin}},\ }\bibfield  {title}
  {\bibinfo {title} {Enormously high concentrations of fluorescent
  nitrogen-vacancy centers fabricated by sintering of detonation
  nanodiamonds},\ }\href {https://doi.org/10.1002/smll.201001887} {\bibfield
  {journal} {\bibinfo  {journal} {Small}\ }\textbf {\bibinfo {volume} {7}},\
  \bibinfo {pages} {1533} (\bibinfo {year} {2011})},\ \Eprint
  {https://arxiv.org/abs/https://onlinelibrary.wiley.com/doi/pdf/10.1002/smll.201001887}
  {https://onlinelibrary.wiley.com/doi/pdf/10.1002/smll.201001887} \BibitemShut
  {NoStop}%
\bibitem [{\citenamefont {Gali}\ \emph {et~al.}(2008)\citenamefont {Gali},
  \citenamefont {Fyta},\ and\ \citenamefont {Kaxiras}}]{Gali2008}%
  \BibitemOpen
  \bibfield  {author} {\bibinfo {author} {\bibfnamefont {A.}~\bibnamefont
  {Gali}}, \bibinfo {author} {\bibfnamefont {M.}~\bibnamefont {Fyta}},\ and\
  \bibinfo {author} {\bibfnamefont {E.}~\bibnamefont {Kaxiras}},\ }\bibfield
  {title} {\bibinfo {title} {Ab initio supercell calculations on
  nitrogen-vacancy center in diamond: Electronic structure and hyperfine
  tensors},\ }\href {https://doi.org/10.1103/PhysRevB.77.155206} {\bibfield
  {journal} {\bibinfo  {journal} {Phys. Rev. B}\ }\textbf {\bibinfo {volume}
  {77}},\ \bibinfo {pages} {155206} (\bibinfo {year} {2008})}\BibitemShut
  {NoStop}%
\bibitem [{\citenamefont {Nizovtsev}\ \emph
  {et~al.}(2010{\natexlab{b}})\citenamefont {Nizovtsev}, \citenamefont {Kilin},
  \citenamefont {Pushkarchuk}, \citenamefont {Pushkarchuk},\ and\ \citenamefont
  {Kuten'}}]{Nizovtsev2010I}%
  \BibitemOpen
  \bibfield  {author} {\bibinfo {author} {\bibfnamefont {A.~P.}\ \bibnamefont
  {Nizovtsev}}, \bibinfo {author} {\bibfnamefont {S.~Y.}\ \bibnamefont
  {Kilin}}, \bibinfo {author} {\bibfnamefont {V.~A.}\ \bibnamefont
  {Pushkarchuk}}, \bibinfo {author} {\bibfnamefont {A.~L.}\ \bibnamefont
  {Pushkarchuk}},\ and\ \bibinfo {author} {\bibfnamefont {S.~A.}\ \bibnamefont
  {Kuten'}},\ }\bibfield  {title} {\bibinfo {title} {Quantum registers based on
  single nv + n13c centers in diamond: I. the spin hamiltonian method},\ }\href
  {https://doi.org/10.1134/S0030400X10020128} {\bibfield  {journal} {\bibinfo
  {journal} {Optics and Spectroscopy}\ }\textbf {\bibinfo {volume} {108}},\
  \bibinfo {pages} {230} (\bibinfo {year} {2010}{\natexlab{b}})}\BibitemShut
  {NoStop}%
\bibitem [{SI()}]{SI}%
  \BibitemOpen
  \href@noop {} {}\bibinfo {note} {See Supplemental Material for the theory of
  the \ce{^{13}C} hyperfine strucutre, experimental details, and additional
  data for \ce{^{13}C} hyperfine structures.}\BibitemShut {Stop}%
\bibitem [{\citenamefont {He}\ \emph {et~al.}(1993)\citenamefont {He},
  \citenamefont {Manson},\ and\ \citenamefont {Fisk}}]{He1993}%
  \BibitemOpen
  \bibfield  {author} {\bibinfo {author} {\bibfnamefont {X.-F.}\ \bibnamefont
  {He}}, \bibinfo {author} {\bibfnamefont {N.~B.}\ \bibnamefont {Manson}},\
  and\ \bibinfo {author} {\bibfnamefont {P.~T.~H.}\ \bibnamefont {Fisk}},\
  }\bibfield  {title} {\bibinfo {title} {Paramagnetic resonance of photoexcited
  n-v defects in diamond. ii. hyperfine interaction with the $^{14}\mathrm{N}$
  nucleus},\ }\href {https://doi.org/10.1103/PhysRevB.47.8816} {\bibfield
  {journal} {\bibinfo  {journal} {Phys. Rev. B}\ }\textbf {\bibinfo {volume}
  {47}},\ \bibinfo {pages} {8816} (\bibinfo {year} {1993})}\BibitemShut
  {NoStop}%
\bibitem [{\citenamefont {Morton}\ and\ \citenamefont
  {Preston}(1978)}]{Morton1978}%
  \BibitemOpen
  \bibfield  {author} {\bibinfo {author} {\bibfnamefont {J.}~\bibnamefont
  {Morton}}\ and\ \bibinfo {author} {\bibfnamefont {K.}~\bibnamefont
  {Preston}},\ }\bibfield  {title} {\bibinfo {title} {Atomic parameters for
  paramagnetic resonance data},\ }\href
  {https://doi.org/https://doi.org/10.1016/0022-2364(78)90284-6} {\bibfield
  {journal} {\bibinfo  {journal} {Journal of Magnetic Resonance (1969)}\
  }\textbf {\bibinfo {volume} {30}},\ \bibinfo {pages} {577 } (\bibinfo {year}
  {1978})}\BibitemShut {NoStop}%
\bibitem [{\citenamefont {Xie}\ \emph {et~al.}(2021)\citenamefont {Xie},
  \citenamefont {Liu}, \citenamefont {Zhang}, \citenamefont {Wong},
  \citenamefont {Zhou}, \citenamefont {Zhao}, \citenamefont {Wang},
  \citenamefont {Lai},\ and\ \citenamefont {Goh}}]{Xie2021}%
  \BibitemOpen
  \bibfield  {author} {\bibinfo {author} {\bibfnamefont {J.}~\bibnamefont
  {Xie}}, \bibinfo {author} {\bibfnamefont {X.}~\bibnamefont {Liu}}, \bibinfo
  {author} {\bibfnamefont {W.}~\bibnamefont {Zhang}}, \bibinfo {author}
  {\bibfnamefont {S.~M.}\ \bibnamefont {Wong}}, \bibinfo {author}
  {\bibfnamefont {X.}~\bibnamefont {Zhou}}, \bibinfo {author} {\bibfnamefont
  {Y.}~\bibnamefont {Zhao}}, \bibinfo {author} {\bibfnamefont {S.}~\bibnamefont
  {Wang}}, \bibinfo {author} {\bibfnamefont {K.~T.}\ \bibnamefont {Lai}},\ and\
  \bibinfo {author} {\bibfnamefont {S.~K.}\ \bibnamefont {Goh}},\ }\bibfield
  {title} {\bibinfo {title} {Fragile pressure-induced magnetism in fese
  superconductors with a thickness reduction},\ }\href
  {https://doi.org/10.1021/acs.nanolett.1c03508} {\bibfield  {journal}
  {\bibinfo  {journal} {Nano Letters}\ }\textbf {\bibinfo {volume} {21}},\
  \bibinfo {pages} {9310} (\bibinfo {year} {2021})},\ \bibinfo {note} {pMID:
  34714653},\ \Eprint
  {https://arxiv.org/abs/https://doi.org/10.1021/acs.nanolett.1c03508}
  {https://doi.org/10.1021/acs.nanolett.1c03508} \BibitemShut {NoStop}%
\bibitem [{\citenamefont {Piermarini}\ \emph {et~al.}(1973)\citenamefont
  {Piermarini}, \citenamefont {Block},\ and\ \citenamefont
  {Barnett}}]{Piermarini1973}%
  \BibitemOpen
  \bibfield  {author} {\bibinfo {author} {\bibfnamefont {G.~J.}\ \bibnamefont
  {Piermarini}}, \bibinfo {author} {\bibfnamefont {S.}~\bibnamefont {Block}},\
  and\ \bibinfo {author} {\bibfnamefont {J.}~\bibnamefont {Barnett}},\
  }\bibfield  {title} {\bibinfo {title} {Hydrostatic limits in liquids and
  solids to 100 kbar},\ }\href {https://doi.org/10.1063/1.1662159} {\bibfield
  {journal} {\bibinfo  {journal} {Journal of Applied Physics}\ }\textbf
  {\bibinfo {volume} {44}},\ \bibinfo {pages} {5377} (\bibinfo {year}
  {1973})},\ \Eprint {https://arxiv.org/abs/https://doi.org/10.1063/1.1662159}
  {https://doi.org/10.1063/1.1662159} \BibitemShut {NoStop}%
\bibitem [{\citenamefont {Angel}\ \emph {et~al.}(2007)\citenamefont {Angel},
  \citenamefont {Bujak}, \citenamefont {Zhao}, \citenamefont {Gatta},\ and\
  \citenamefont {Jacobsen}}]{Angel2007}%
  \BibitemOpen
  \bibfield  {author} {\bibinfo {author} {\bibfnamefont {R.~J.}\ \bibnamefont
  {Angel}}, \bibinfo {author} {\bibfnamefont {M.}~\bibnamefont {Bujak}},
  \bibinfo {author} {\bibfnamefont {J.}~\bibnamefont {Zhao}}, \bibinfo {author}
  {\bibfnamefont {G.~D.}\ \bibnamefont {Gatta}},\ and\ \bibinfo {author}
  {\bibfnamefont {S.~D.}\ \bibnamefont {Jacobsen}},\ }\bibfield  {title}
  {\bibinfo {title} {Effective hydrostatic limits of pressure media for
  high-pressure crystallographic studies},\ }\href
  {https://doi.org/10.1107/S0021889806045523} {\bibfield  {journal} {\bibinfo
  {journal} {Journal of Applied Crystallography}\ }\textbf {\bibinfo {volume}
  {40}},\ \bibinfo {pages} {26} (\bibinfo {year} {2007})},\ \Eprint
  {https://arxiv.org/abs/https://onlinelibrary.wiley.com/doi/pdf/10.1107/S0021889806045523}
  {https://onlinelibrary.wiley.com/doi/pdf/10.1107/S0021889806045523}
  \BibitemShut {NoStop}%
\bibitem [{\citenamefont {Tateiwa}\ and\ \citenamefont
  {Haga}(2009)}]{Tateiwa2009}%
  \BibitemOpen
  \bibfield  {author} {\bibinfo {author} {\bibfnamefont {N.}~\bibnamefont
  {Tateiwa}}\ and\ \bibinfo {author} {\bibfnamefont {Y.}~\bibnamefont {Haga}},\
  }\bibfield  {title} {\bibinfo {title} {Evaluations of pressure-transmitting
  media for cryogenic experiments with diamond anvil cell},\ }\href
  {https://doi.org/10.1063/1.3265992} {\bibfield  {journal} {\bibinfo
  {journal} {Review of Scientific Instruments}\ }\textbf {\bibinfo {volume}
  {80}},\ \bibinfo {pages} {123901} (\bibinfo {year} {2009})},\ \Eprint
  {https://arxiv.org/abs/https://doi.org/10.1063/1.3265992}
  {https://doi.org/10.1063/1.3265992} \BibitemShut {NoStop}%
\bibitem [{\citenamefont {Klotz}\ \emph {et~al.}(2009)\citenamefont {Klotz},
  \citenamefont {Chervin}, \citenamefont {Munsch},\ and\ \citenamefont
  {Marchand}}]{Klotz2009}%
  \BibitemOpen
  \bibfield  {author} {\bibinfo {author} {\bibfnamefont {S.}~\bibnamefont
  {Klotz}}, \bibinfo {author} {\bibfnamefont {J.-C.}\ \bibnamefont {Chervin}},
  \bibinfo {author} {\bibfnamefont {P.}~\bibnamefont {Munsch}},\ and\ \bibinfo
  {author} {\bibfnamefont {G.~L.}\ \bibnamefont {Marchand}},\ }\bibfield
  {title} {\bibinfo {title} {Hydrostatic limits of 11 pressure transmitting
  media},\ }\href {https://doi.org/10.1088/0022-3727/42/7/075413} {\bibfield
  {journal} {\bibinfo  {journal} {Journal of Physics D: Applied Physics}\
  }\textbf {\bibinfo {volume} {42}},\ \bibinfo {pages} {075413} (\bibinfo
  {year} {2009})}\BibitemShut {NoStop}%
\bibitem [{\citenamefont {Manson}\ \emph {et~al.}(1990)\citenamefont {Manson},
  \citenamefont {He},\ and\ \citenamefont {Fisk}}]{Manson1990}%
  \BibitemOpen
  \bibfield  {author} {\bibinfo {author} {\bibfnamefont {N.~B.}\ \bibnamefont
  {Manson}}, \bibinfo {author} {\bibfnamefont {X.-F.}\ \bibnamefont {He}},\
  and\ \bibinfo {author} {\bibfnamefont {P.~T.~H.}\ \bibnamefont {Fisk}},\
  }\bibfield  {title} {\bibinfo {title} {Raman heterodyne detected
  electron-nuclear-double-resonance measurements of the nitrogen-vacancy center
  in diamond},\ }\href {https://doi.org/10.1364/OL.15.001094} {\bibfield
  {journal} {\bibinfo  {journal} {Opt. Lett.}\ }\textbf {\bibinfo {volume}
  {15}},\ \bibinfo {pages} {1094} (\bibinfo {year} {1990})}\BibitemShut
  {NoStop}%
\bibitem [{\citenamefont {Sz\'asz}\ \emph {et~al.}(2013)\citenamefont
  {Sz\'asz}, \citenamefont {Hornos}, \citenamefont {Marsman},\ and\
  \citenamefont {Gali}}]{Szasz2013Hyperfine}%
  \BibitemOpen
  \bibfield  {author} {\bibinfo {author} {\bibfnamefont {K.}~\bibnamefont
  {Sz\'asz}}, \bibinfo {author} {\bibfnamefont {T.}~\bibnamefont {Hornos}},
  \bibinfo {author} {\bibfnamefont {M.}~\bibnamefont {Marsman}},\ and\ \bibinfo
  {author} {\bibfnamefont {A.}~\bibnamefont {Gali}},\ }\bibfield  {title}
  {\bibinfo {title} {Hyperfine coupling of point defects in semiconductors by
  hybrid density functional calculations: The role of core spin polarization},\
  }\href {https://doi.org/10.1103/PhysRevB.88.075202} {\bibfield  {journal}
  {\bibinfo  {journal} {Phys. Rev. B}\ }\textbf {\bibinfo {volume} {88}},\
  \bibinfo {pages} {075202} (\bibinfo {year} {2013})}\BibitemShut {NoStop}%
\bibitem [{\citenamefont {Swift}\ \emph {et~al.}(2020)\citenamefont {Swift},
  \citenamefont {Peelaers}, \citenamefont {Mu}, \citenamefont {Morton},\ and\
  \citenamefont {Van~de Walle}}]{Swift2020First}%
  \BibitemOpen
  \bibfield  {author} {\bibinfo {author} {\bibfnamefont {M.~W.}\ \bibnamefont
  {Swift}}, \bibinfo {author} {\bibfnamefont {H.}~\bibnamefont {Peelaers}},
  \bibinfo {author} {\bibfnamefont {S.}~\bibnamefont {Mu}}, \bibinfo {author}
  {\bibfnamefont {J.~J.~L.}\ \bibnamefont {Morton}},\ and\ \bibinfo {author}
  {\bibfnamefont {C.~G.}\ \bibnamefont {Van~de Walle}},\ }\bibfield  {title}
  {\bibinfo {title} {First-principles calculations of hyperfine interaction,
  binding energy, and quadrupole coupling for shallow donors in silicon},\
  }\href {https://doi.org/10.1038/s41524-020-00448-7} {\bibfield  {journal}
  {\bibinfo  {journal} {npj Computational Materials}\ }\textbf {\bibinfo
  {volume} {6}},\ \bibinfo {pages} {181} (\bibinfo {year} {2020})}\BibitemShut
  {NoStop}%
\end{thebibliography}%

\end{document}